\documentstyle{article}
\newcommand{\be}{\begin{equation}}
\newcommand{\ee}{\end{equation}}
\newcommand{\bea}{\begin{array}}
\newcommand{\eea}{\end{array}}

\title{THE WHITHAM EQUATIONS REVISITED}
\author{ROBERT CARROLL and JEN HSU CHANG\\
University of Illinois\\
Urbana, IL 61801\thanks{email:  rcarroll@symcom.math.uiuc.edu}}

\date{November, 1996}

\begin{document}
\bibliographystyle{plain}
\maketitle

\begin{abstract} 

We survey some topics involving the Whitham equations, concentrating on
the role of $\psi\psi^*$ (or square eigenfunctions) in averaging.
Some connections to symplectic geometry and Seiberg-Witten theory are
indicated.

\end{abstract}


\section{INTRODUCTION}
\renewcommand{\theequation}{1.\arabic{equation}}\setcounter{equation}{0}

A preliminary sketch of some of this material was given in \cite{cn} and
in view of several recent developments (cf. \cite{bz,co,cp,cs,dq,ib,
kj,kt,kf,my,mz,ne})
we decided to expand and rewrite \cite{cn} in two parts, of which this
is the first.  The second part will deal with theory of kernels related
to $\psi\psi^*$ (cf. \cite{cq}).
\\[3mm]\indent
There are many physical situations where averaging procedures of various
types are employed and we refer to \cite{wa} for a general background
(see \cite{da,db,dd,de,fa,fd,fb,ka,kc,ke,ki,
mt,mc} for weakly deformed soliton lattices, multiphase averaging,
etc.).  The Whitham equations arise in this context and we will look
at some of this here; we attempt to organize a subset of information in a
coherent and
more or less rigorous manner, while clarifying various
formulas and assertions in the literature.  Connections to
dispersionless theory are indicated and material has been
selected which will be useful in further studies of
topological field theory (TFT),
Landau-Ginzburg (LG) theory, Seiberg-Witten (SW) theory, 
etc.  The exposition in this first paper
is mainly expository but we indicate
some new connections and results (cf. also \cite{co,cp,cq,cs}).
The principal theme involves averaging of square
eigenfunctions for KdV or averaging of quantities related to
$\psi\psi^*$ for KP, where $\psi$ is the Baker-Akhiezer (BA) function;
this point of view (along with general use of the BA function) is emphasized
for derivation of Whitham equations, construction of differentials, and 
moduli considerations.
Many basic formulas
involving theta functions, differentials, etc. are also exhibited in order
to make the text useful as a launching pad for further investigation.

\section{CURVES AND MODULI}
\renewcommand{\theequation}{2.\arabic{equation}}\setcounter{equation}{0}

\subsection{Riemann surfaces and BA functions}

Following \cite{bc,bb,ca,cm,dc,dd,fz,ge,kt,tb} we
take an arbitrary Riemann surface $\Sigma$
of genus $g$, pick a point $Q$ and a local variable $1/k$ near $Q$
such that $k(Q) = \infty$, and, for illustration, take $q(k)=kx+k^2y+k^3t$. 
Let $D = P_1 + \cdots + P_g$ be a non-special
divisor of degree $g$ and write $\psi$ for the (unique up to a constant
multiplier by virtue of the Riemann-Roch theorem)
Baker-Akhiezer (BA) function characterized by the properties
({\bf A}) $\psi$ is meromorphic on $\Sigma$ except for $Q$ where $\psi
(P)exp(-q(k))$ is analytic 
and (*) $\psi\sim exp(q(k))[1 + \sum_1^{\infty}(\xi_j/
k^j)]$ near $Q$.  ({\bf B}) On $\Sigma/Q,\,\,\psi$ has only a finite
number of poles (at the $P_i$).
In fact $\psi$ can be taken in the form ($P\in\Sigma,\,\,
P_0\not= Q$)
\be
\psi(x,y,t,P) =  exp[\int^P_{P_0}(xd\Omega^1 + yd\Omega^2 + td\Omega^3)]
\cdot\frac{\Theta({\cal A}(P) + xU + yV + tW + z_0)}{\Theta({\cal A}
(P) + z_0)}
\label{psi}
\ee
where $d\Omega^1 = dk + \cdots,\,\,d\Omega^2 = d(k^2) + \cdots,\,\,
d\Omega^3 = d(k^3) + \cdots, U_j = \int_{B_j}d\Omega^1,\,\,V_j = \int_
{B_j}d\Omega^2,\,\,W_j = \int_{B_j}d\Omega^3\,\,(j = 1,\cdots,g),\,\,z_0
= -{\cal A}(D) - K$, and $\Theta$ is the Riemann theta function.
The symbol $\sim$ will be used generally to mean ``corresponds
to" or ``is associated with"; occasionally it also denotes asymptotic
behavior; this should be clear from the context.
Here the
$d\Omega_j$ are meromorphic differentials of second kind normalized via
$\int_{A_k}d\Omega_j = 0\,\,(A_j,\,B_j$ are canonical homology cycles)
and we note that $xd\Omega^1 + yd\Omega^2 + td\Omega^3\sim
dq(k)$ normalized.
${\cal A}\sim$ Abel-Jacobi map ${\cal A}(P) = (\int^P_{P_0}d\omega_k)$
where
the $d\omega_k$ are normalized holomorphic differentials, $k = 1,\cdots,g,
\,\,\int_{A_j}d\omega_k = \delta_{jk}$, and $K = (K_j)\sim$ Riemann
constants ($2K = -{\cal A}(K_{\Sigma})$ where $K_{\Sigma}$ is the
canonical class of $\Sigma\sim$ equivalence class of meromorphic 
differentials).  Thus $\Theta({\cal A}(P) + z_0)$ has exactly $g$ zeros
(or vanishes identically.  The paths of integration are to be the
same in computing $\int_{P_0}^Pd\Omega^i$ or ${\cal A}(P)$ and it is
shown in \cite{bc,bb,cn,dc} that $\psi$ is well defined (i.e. path
independent).  
Then the $\xi_j$ in (*) can be computed
formally and one determines Lax operators $L$ and $A$ such that
$\partial_y\psi = L\psi$ with $\partial_t\psi = A\psi$.  Indeed, given
the $\xi_j$ write $u = -2\partial_x\xi_1$ with $w = 3\xi_1\partial_x\xi_1
-3\partial^2_x\xi_1 - 3\partial_x\xi_2$.  Then formally, near $Q$, one
has $(-\partial_y + \partial_x^2 + u)\psi = O(1/k)exp(q)$ and 
$(-\partial_t + \partial^3_x + (3/2)u\partial_x + w)\psi = O(1/k)exp(q)$
(i.e. this choice of $u,\,w$ makes the coefficients of $k^nexp(q)$ vanish
for $n = 0,1,2,3$).  Now define $L = \partial_x^2 + u$ and $A = \partial^3_x
+ (3/2)u\partial_x + w$ so $\partial_y\psi = L\psi$ and $\partial_t\psi
= A\psi$.  This follows from the uniqueness of BA functions with the same
essential singularity and pole divisors (Riemann-Roch). 
Then we have, via compatibility
$L_t - A_y = [A,L]$, a KP equation $(3/4)u_{yy} = \partial_x[u_t
-(1/4)(6uu_x + u_{xxx})]$ and therefore such KP equations are parametrized
by nonspecial divisors or equivalently by points in general position
on the Jacobian variety $J(\Sigma)$.
The flow variables $x,y,t$ are put in by hand in ({\bf A}) via
$q(k)$ and then miraculously reappear in the theta function via
$xU+yV+tW$; thus the Riemann surface itself contributes to establish
these as linear flow variables on the Jacobian.
The pole positions
$P_i$ do not vary with $x,y,t$ and $(\dagger)\,\,
u = 2\partial^2_x log\Theta(xU + yV  + tW + z_0) + c$ 
exhibits $\Theta$ as a tau function.
\\[3mm]\indent
We recall also that
a divisor $D^{*}$ of degree $g$ is dual to $D$ (relative to $Q$) if
$D + D^{*}$ is the null divisor of a meromorphic differential $d\hat{\Omega}
= dk + (\beta/k^2)dk + \cdots$ with a double pole at $Q$ (look at
$\zeta = 1/k$ to recognize the double pole).  Thus 
$D + D^{*} -2Q\sim K_{\Sigma}$ so ${\cal A}(D^{*}) - {\cal A}(Q) + K =
-[{\cal A}(D) - {\cal A}(Q) + K]$.  One can define then a function
$\psi^{*}(x,y,t,P) = exp(-kx-k^2y-k^3t)[1 + \xi_1^{*}/k) + \cdots]$
based on $D^{*}$ (dual BA function)
and a differential $d\hat{\Omega}$ with zero divisor $D+D^*$, such that
$\phi = \psi\psi^{*}d\hat{\Omega}$ is
meromorphic, having for poles only
a double pole at $Q$ (the zeros of $d\hat{\Omega}$ cancel
the poles of $\psi\psi^{*}$).  Thus $\psi\psi^*d\hat{\Omega}\sim \psi\psi^*(1+
(\beta/k^2+\cdots)dk$ is meromorphic with a second order pole at $\infty$,
and no other poles.  
For $L^{*} = L$ and $A^{*} = -A + 2w
-(3/2)u_x$ one has then $(\partial_y + L^{*})\psi^{*} = 0$ and
$(\partial_t + A^{*})\psi^{*} = 0$.  Note that 
the prescription above seems to specify for $\psi^*$
($\vec{U}=xU+yV+tW,\,\,z_0^* =
-{\cal A}(D^*)-K$)
\be
\psi^*\sim e^{-\int^P_{P_o}(xd\Omega^1+yd\Omega^2+td\Omega^3)}
\cdot\frac{\Theta({\cal A}(P)-\vec{U}+z_0^*)}
{\Theta({\cal A}(P)+z_0^*)}
\label{star}
\ee
\indent
In any event the message here is that for any Riemann surface 
$\Sigma$ one can
produce a BA function $\psi$ with assigned flow variables $x,y,t,\cdots$
and this $\psi$ gives rise to a 
(nonlinear) KP equation with solution $u$ linearized
on the Jacobian $J(\Sigma)$.
For averaging with KP (cf. \cite{cn,fb,ka}) 
we can use formulas (cf. (\ref{psi}) and
(\ref{star}))
\be
\psi = e^{px+Ey+\Omega t}\cdot\phi(Ux+Vy
+Wt,P)
\label{YDD}
\ee
\be
\psi^* = e^{-px-Ey-\Omega t}\cdot\phi^*
(-Ux-Vy-Wt,P)
\label{YEE}
\ee
to isolate the quantities of interest in averaging
(here $p=p(P),\,\,E = E(P),\,\,\Omega =
\Omega(P),$ etc.)
We think here of a general Riemann surface $\Sigma_g$ with holomorphic
differentials $d\omega_k$ and quasi-momenta and quasi-energies
of the form $dp=d\Omega^1,\,\,dE=d\Omega^2,\,\,d\Omega=d\Omega^3,\cdots
\,\,(p=\int_{P_0}^Pd\Omega^1$ etc.) where the $d\Omega^j=d\Omega_j=
d(\lambda^j+O(\lambda^{-1}))$ are meromorphic differentials of the second
kind.  Following \cite{ka} one could
normalize now via $\Re\int_{A_j}d\Omega^k=
\Re\int_{B_j}d\Omega^k=0$.  Then write e.g. 
$U_k=(1/2\pi i)\oint_{A_k}dp$ and $U_{k+g}=-(1/2\pi i)\oint_{B_k}dp\,\,
(k=1,\cdots,g)$ with similar stipulations for $V_k\sim\oint d\Omega^2,\,\,
W_k\sim\oint d\Omega^3,$ etc.  This leads to real $2g$ period vectors
and evidently one could also normalize via $\oint_{A_m}d\Omega^k=0$ or
$\Im\oint_{A_m}d\Omega^k=\Im\oint_{B_m}d\Omega^k=0$ (further we set
$B_{jk}=\oint_{B_k}d\omega_j$).

\subsection{General remarks on averaging}

Averaging can be rather mysterious at first due to some hasty treatments
and bad choices of notation - plus many inherent difficulties.  Some
of the clearest exposition seems to be in \cite{ba,fa,fd,mc,wa}
whereas the more extensive developments in e.g. 
\cite{da,db,dd,de,ka,ke} may be confusing at first,
until one realizes what is going on. 
\\[3mm]\indent
One choice of background situation
involves an examination of dispersionless limits with no concern for
periodicity or Riemann surfaces (cf. here 
\cite{ae,ch,ci,cl,gh,kn,ta}).  Thus given
e.g. a KdV equation 
$u_t + 6uu_x + u_{xxx} = 0$ set $\epsilon x = X$ and $\epsilon t = T$
leading to $u^{\epsilon}_T + 6u^{\epsilon}u^{\epsilon}_X + 
\epsilon^2 u^{\epsilon}_{XXX} = 0$.  The study of $u^{\epsilon}\to
\hat{u}$ where $\hat{u}$ satisfies the Euler equation $\hat{u}_T + 6
\hat{u}\hat{u}_X = 0$ is a very delicate matter, of great interest in
applications and in PDE but we do not discuss this here (cf. \cite{la,vc}).
On the other hand the purely algebraic
passage of the background mathematics of KdV (involving Lax operators,
the KdV hierarchy, tau functions, vertex operators, etc.) to the 
corresponding background mathematics of the dispersionless theory, is
relatively easy and will be indicated below.
Moreover it is of great importance in an entirely 
different direction, namely in the study of 
topological field theory, strings, and 2-D gravity (see e.g. 
\cite{ab,ae,ch,ci,cl,cr,dl,di,dj,dk,kc,kj,kl,vb}). 
We will insert such material later as appropriate. 
\\[3mm]\indent
Let us follow \cite{kc} (cf. also \cite{kb}) in order to have a 
suitably complicated example leading directly to matters of interest here,
so as background we consider the KP equation
\be
\frac{3}{4}\sigma^2u_{yy} = \partial_x[u_t -\frac{3}{2}uu_x +
\frac{1}{4}u_{xxx}]
\label{AC}
\ee
($\sigma^2 = 1\sim$ KP-2; $\sigma^2 = -1\sim$ KP-1).  Here $\sigma L
\sim \partial^2_x - u$ with $\sigma\partial_y\psi = 
(\partial^2_x - u)\psi$ and $A\sim
\partial^3_x - (3/2)u\partial_x + w$, which is slightly different than
before (to connect notations let $u\to -u$ and insert $\sigma$).  Then
one has the compatibility equations $[\partial_y - L,\partial_t - A] = 0$
as before and it should be noted that for any function $g(t),\,\,t\sim
x,y,t,t_4,\cdots$, operators $\tilde{L} = gLg^{-1} + \partial_yg\,g^{-1}$
and $\tilde{A} = gAg^{-1} + \partial_t g\,g^{-1}$ could be used,
corresponding to a new wave function $\tilde{\psi} = g\psi$.  The notation
of \cite{kc} also involves differentials $d\Omega_i\sim d\Omega^i$ where
$d\Omega_i = dk^i(1 + O(1/k)]$ and $\int_{A_k}d\Omega_j = 0$ with
$U^k_j = (1/2\pi i)\int_{B_k}d\Omega_j$ (so $U^k_1\sim U_k,\,\,U^k_2\sim
V_k$, and $U^k_3\sim W_k$, up to factors of $2\pi i$).  There will be solutions
$\psi$ as before in (\ref{psi}) with potentials $u$ given via $(\dagger)$.
We recall that Riemann theta functions are more precisely written as
$\Theta(z|B)$ where $B\sim$ a certain matrix (cf. below),
and $z\sim (z_1,\cdots,z_g)
\sim xU+yV+tW+z_0$ for example.  One knows via \cite{ca,dg} that
$u$ of the form $(\dagger)\,\,u = 2\partial^2_xlog\Theta(z|B) + c$ is a 
solution of KP if and only if the matrix $B$ defining $\Theta$ is the
B-period matrix of a Riemann surface determined via $\int_{B_k}d\omega_j$
(with $z$ as defined).
Now consider the spectral theory of 2-D periodic operators (**) $(\sigma
\partial_y - \partial_x^2 + u(x,y))\psi = 0$ where $u(x,y) =
u(x+a_1,y) = u(x,y+a_2)$.  Bloch solutions are defined via
\be
\psi(x+a_1,y,w_1,w_2) =  w_1\psi(x,y,w_1,w_2);
\label{AD}
\ee
$$\psi(x,y+a_2,w_1,w_2) = w_2\psi(x,y,w_1,w_2)$$
and we assume $\psi(0,0,w_1,w_2) = 1$.  The pairs $Q=(w_1,w_2)$ for which
there exists such solutions is called the Floquet set $\Gamma$ and the
multivalued functions $p(Q)$ and $E(Q)$ such that $w_1 = exp(ipa_1)$ with
$w_2 = exp(iEa_2)$ are called quasi-momentum and quasi-energy respectively.
By a gauge transformation $\psi\to exp(h(y))\psi$, with $\partial_y
h(y)$ periodic one obtains solutions of (**) with a new potential
$\tilde{u} = u - \sigma\partial_y h$ so we can assume $\int_0^{a_1}
u(x,y)dx = 0$.
\\[3mm]\indent
For $M_0 = \sigma\partial_y
-\partial_x^2$ (with $u=0$) the Floquet set is parametrized by $k\in
{\bf C}$ such that $w_1^0 = exp(ika_1)$ and $w_2^0 = exp(-k^2a_2/\sigma)$
and the Bloch solutions are $\psi(x,y,k) = exp(ikx-k^2y/\sigma)$.
The adjoint Bloch solutions are $\psi^{+}(x,y,k) = exp(-ikx+k^2y/\sigma)$
satisfying $(\sigma\partial_y + \partial_x^2)\psi^{+} = 0$.  The image
of the map $k\to (w_1^0,w_2^0)\in{\bf C}^2$ is the Floquet set for $M_0$
corresponding to the Riemann surface with intersections corresponding
to pairs $k\not= k'$ such that $w_i^0(k) = w_i^0(k'),\,\,i = 1,2$.
This means $k-k' = (2\pi N/a_1)$ and $k^2-(k')^2 = (2\pi i\sigma M/a_2)$
where $N,\,M$ are integers, creating "resonant" points $k = k_{N,M} =
(\pi N/a_1) - (i\sigma Ma_1/Na_2),\,\,N\not= 0,\,\,k' = k_{-N,-M}$.
Then for $k_0\not=k_{N,M}$ and $u$ sufficiently small
one can construct a formal Bloch solution 
of (**) in the form of a convergent perturbation series (for any
$\sigma$).  Thus outside of some neighborhoods of the resonant points
one can obtain a Bloch solution $\tilde{\psi}(x,y,k_0)$ which is
analytic in $k_0$, but the extension of $\tilde{\psi}$ to a resonant
domain can be tricky.  For $\Re\sigma = 0$ (as in KP-1)
the resonant points are dense
on the real axis whereas for $\Re\sigma\not= 0$ (as in KP-2) 
there are only a finite
number of resonant points in any finite domain of {\bf C}.  In the latter
case one can glue handles between domains around resonant points
and create a Riemann surface $\Gamma$
of Bloch solutions $\psi(x,y,Q),\,\,Q\in\Gamma$. 
Moreover, if the potential
$u$ can be analytically extended into a domain $|\Im x|<\tau_1,\,\,
|\Im y|<\tau_2$, then the technique can also
be adapted even when $u$ is not small.
Such Bloch solutions, normalized by $\psi(0,0,Q) = 1$, are meromorphic
on $\Gamma$ and in the case of a finite number of handles a one point
compactification of $\Gamma$ is obtained so that $\psi$ is in fact the
BA function for $\Gamma$.  Generally speaking finite zone situations
as in (\ref{psi}) with potentials given by Riemann 
theta functions are quasi-periodic in nature.  To single out 
conditions for periodicity one asks for meromorphic differentials
$dp$ and $dE$ on $\Gamma$ having their only singularities at $Q\sim$
point at $\infty$ of the form $dp = dk(1+O(k^{-2}))$ and $dE = i\sigma^{-1}
dk^2(1 + O(k^{-3}))$, normalized so that all periods are real, and 
satisfying, for any cycle $C$ on $\Gamma$, $\oint_Cdp = (2\pi n_C/a_1)$
with $\oint_CdE = (2\pi m_C/a_2)$ where $n_C,\,m_C$ are integers.
Then the corresponding potentials $u(x,y)$ will have periods $a_1$ and
$a_2$ in $x$ and $y$, with multipliers $w_1(P) = exp(ia_1\int^P dp)$
and $w_2(P) = exp(ia_2\int^P dE)$.
We go now to finite zone (or quasiperiodic) situations as in (\ref{psi}) 
with potentials as in $(\dagger)$, where $z\sim xU + yV + tW
+ z_0$ can be written as $(z_k) = [x\int_{b_k}\Omega^1 + y\int_{b_k}
\Omega^2 + t\int_{b_k}\Omega^3 + z^0_k] = (\zeta_k + z^0_k),\,\,
k = 1,\cdots,g$.  Since $\Theta(z+2\pi iN) = \Theta(z)$ we could set
$\zeta = i\theta$ so that as a function of $\theta = (\theta_k),\,\,u$ is
periodic of period $2\pi$ in each variable $\theta_k$.  Now one wants to
consider modulated finite zone situations where solutions are of the form
$u = u_0(xU+yV+tW|I) = u_0(\theta_1,\cdots,\theta_g|I_1,\cdots,I_n)$ where
$u_0$ is periodic in the $\theta_j$ with $U,V,W = U,V,W(I)$.  One assumes
there will be slow variables $X = \epsilon x,\,\,Y = \epsilon y,$ and
$T = \epsilon t$ with fast variables $x,y,t$ so that "asymptotic" solutions
\be
u = u_0(\frac{1}{\epsilon}\hat{S}(X,Y,T)|I(X,Y,T)) + \epsilon u_1(x,y,t) +
\epsilon^2u_2(x,y,t) + \cdots
\label{AE}
\ee
can be envisioned.  In practice the parameters $I_k$ depend on the
moduli of our Riemann surface, which are then allowed to change
smoothly with the slow variables $X,Y,T$ (so $U,V,W$ also can change),
and one looks for solutions (\ref{AE}) with uniformly bounded $u_1$.
We do not look for convergence of the series in (\ref{AE}) nor check
any other features of ``asymptotic solution" or ``asymptotic series".
Such procedures are standard in the study
of what are called weakly deformed soliton
lattices and we emphasize that the introduction of slow variables
here is an assumption based on modulated wave trains.

\subsection{Hyperelliptic curves and KdV averaging}

One knows that hyperelliptic curves play a special role in the theory
of algebraic curves and Riemann surfaces (see e.g. \cite{bc,bb,ca,cn,cm,fg,fe,
fa,fb,ga,ma,nd,sb}).  For 
hyperelliptic Riemann surfaces one can pick
any $2g+2$ points $\lambda_j\in {\bf P}^1$ and there will be a unique
hyperelliptic curve $\Sigma_g$ with a 2-fold map $f:\,\Sigma_g\to{\bf P}^1$ 
having branch locus $B=\{\lambda_j\}$.  Since any 3 points $\lambda_i,\,
\lambda_j,\,\lambda_k$ can be sent to $0,\,1,\,\infty$ by an automorphism
of ${\bf P}^1$ the general hyperelliptic surface of genus $g$ can be
described by $(2g+2)-3=2g-1$ points on ${\bf P}^1$.  Since $f$ is unique
up to an automorphism of ${\bf P}^1$ any hyperelliptic $\Sigma_g$ corresponds
to only finitely many such collections of $2g-1$ points so locally there
are $2g-1$ (moduli) parameters.  Since the moduli space of algebraic
curves has dimension $3g-3$ one sees that for $g\geq 3$ the generic 
Riemann surface is nonhyperelliptic whereas for $g=2$ all Riemann
surfaces are hyperelliptic (with 3 moduli).  For $g=1$ we have tori
or elliptic curves with one modulus $\tau$ and $g=0$ corresponds to
${\bf P}^1$.  In many papers on soliton mathematics and integrable systems
one takes real distinct branch points $\lambda_j,\,\,1\leq j\leq 2g+1$, and 
$\infty$, with $\lambda_1<\lambda_2<\cdots<\lambda_{2g+1}<\infty$ and
\be
\mu^2=\prod_1^{2g+1}(\lambda-\lambda_j)=P_{2g+1}(\lambda,\lambda_j)
\label{mu}
\ee
as the defining equation for $\Sigma_g$.  Evidently one could choose
$\lambda_1=0,\,\,\lambda_2=1$ in addition so for $g=1$ we could use
$0<1<u<\infty$ for a familiar parametrization with elliptic integrals,
etc.  One can take $d\lambda/\mu,\,\,\lambda d\lambda/\mu,\cdots,
\lambda^{g-1} d\lambda/\mu$ as a basis of holomorphic differentials on
$\Sigma_g$ but usually one takes linear combinations of these denoted
by $d\omega_j,\,\,1\leq j\leq g$, normalized via $\oint_{A_i}d\omega_j=
\delta_{ij}$, with period matrix defined via $\oint_{B_i}d\omega_j=
\Pi_{ij}$.  The matrix $\Pi=(\Pi_{ij})$ is symmetric with $\Im\Pi>0$ and
it determines the curve.  Frequently in situations arising from KdV
(Korteweg-deVries equation) for example one regards the intervals
$[\lambda_1,\lambda_2],\cdots,[\lambda_{2g+1},\infty)$ as spectral
bands and intervals $(\lambda_2,\lambda_3),\cdots,(\lambda_{2g},\lambda_
{2g+1})$ as gaps with the $a_i$ cycles around $(\lambda_{2i},\lambda_
{2i+1})\,\,(i=1,\cdots,g)$.
One will also want to consider another representation of hyperelliptic
curves of genus $g$ via
\be
\mu^2=\prod_0^{2g+1}(\lambda-\lambda_j)=P_{2g+2}(\lambda,\lambda_j)
\label{muu}
\ee
where $\infty$ is now not a branch point and in fact there are two points
$\mu_{\pm}$ corresponding to $\lambda=\infty$.
\\[3mm]\indent
We recall next some of the results and techniques of \cite{fa,mc}
where one can see explicitly the nature of things.
The presentation here follows \cite{cn}.
First from \cite{mc}, in a slightly different notation,
write $q_t = 6qq_x - q_{xxx}$ with Lax pair
$L = -\partial^2_x + q,\,\,B = -4\partial^3_x + 3(q\partial_x + \partial_x q),
\,\,L_t = [B,L],\,\,L\psi = \lambda\psi,$ and $\psi_t = B\psi$.
Let $\psi$ and $\phi$ be two solutions of the Lax pair equations and
set $\Psi = \psi\phi$; these are the very important ``square eigenfunctions"
which arise in many ways with interesting and varied meanings 
(cf. \cite{cd,ce,ch,ci,cl}).  Evidently
$\Psi$ satisfies
\be
[-\partial^3_x + 2(q\partial_x+\partial_x q)]\Psi = 4\lambda\partial_x\Psi;
\,\,\partial_t\Psi = -2q_x\Psi + 2(q+2\lambda)\partial_x\Psi
\label{AF}
\ee
From (\ref{AF}) one finds immediately the conservation law ({\bf C}):
$\partial_t[\Psi] + \partial_x[6(q-2\lambda)\Psi - 2\partial^2_x\Psi] = 0$.
If one looks for solutions of (\ref{AF}) of the form $\Psi(x,t,\lambda) =
1 + \sum_1^{\infty}[\Psi_j(x,t)]\lambda^{-j}$ as $\lambda\to\infty$ then
one obtains a recursion relation for polynomial densities
\be
\partial_x\Psi_{j+1} = [-\frac{1}{2}\partial^3_x + (q\partial_x + 
\partial_x q)]\Psi_j\,\,(j = 1,2,\cdots);\,\,\Psi_0 = 1
\label{AG}
\ee
Now consider the operator $L = -\partial^2_x + q$ in $L^2(-\infty,\infty)$
with spectrum consisting of closed intervals separated by exactly $N$ gaps
in the spectrum.  The $2N+1$ endpoints $\lambda_k$
of these spectral bands are
denoted by $-\infty<\lambda_0<\lambda_1<\cdots<\lambda_{2N}<\infty$ 
and called the
simple spectrum of $L$.  They can be viewed as constants of motion for
KdV when $L$ has this form.  We are dealing here with the hyperelliptic
Riemann surface determined via $R^2(\lambda)=\prod_0^{2N}(\lambda - \lambda_k)$
($\infty$ is a branch point)
and one can think of a manifold ${\cal M}$ of $N$-phase waves with
fixed simple spectrum as an $N$-torus based on $\theta_j\in [0,2\pi)$.
Hamiltonians in the KdV hierarchy generate flows on this torus and one
writes $q = q_N(\theta_1,\cdots,\theta_N)$ (the $\theta$ variables
originate as in our previous discussion if we use theta functions
for the integration - cf. also below).  
Now there is no $y$ variable so let us write
$\theta_j = x\kappa_j+tw_j$ (we will continue to use $d\omega_j$ for
normalized holomorphic differentials)
For details concerning the Riemann surface we refer to \cite{bb,ca,dc,fa,nd} and
will summarize here as follows.  For any $q_N$ as indicated one can find
functions $\mu_j(x,t)$ via $\Psi(x,t,\lambda) = \prod_1^N(\lambda -
\mu_j(x,t))$ where $\mu_j(x,t)\in [\lambda_{2j-1},\lambda_{2j}]$ and
satisfies
\be
\partial_x\mu_j = -2i(R(\mu_j)/\prod_{i\not= j}(\mu_j-\mu_i));
\label{AH}
\ee
$$\partial_t\mu_j = -2i[2(\sum_0^{2N}\lambda_k - 2\sum_{i\not= j}\mu_i)]
\cdot (R(\mu_j)/\prod_{i\not= j}(\mu_j - \mu_i)$$
In fact the $\mu_j$ live on the Riemann surface of $R(\lambda)$
in the spectral gaps
and as $x$ increases $\mu_j$ travels from $\lambda_{2j-1}$ to $\lambda_{2j}$
on one sheet and then returns to $\lambda_{2j-1}$ on the other sheet; this
path will be called the $j^{th}\,\mu$-cycle ($\sim A_j$).  
In the present context we
will write the theta function used for integration purposes as
$\Theta({\bf z},\tau) = \sum_{m\in {\bf Z}^N}exp[\pi i(2({\bf m},{\bf z})
+ ({\bf m},\tau{\bf m}))]$ where ${\bf z}\in {\bf Z}^N$ and $\tau$ 
denotes the $N\times N$ period matrix ($\tau$ is symmetric with $\Im\tau
> 0$).  We take canonical cuts $A_i,\,B_i\,\,(i = 1,\cdots,N)$ and
let $d\omega_j$ be holomorphic diffentials normalized via
$\int_{a_j} d\omega_k = \delta_{jk}$ (the cycle $A_j$ corresponds to
a loop around the cut $A_j$).  Then $q_N$ can be represented in the form
\be
q_N(x,t) = \Lambda + \Gamma -2\partial^2_xlog\Theta({\bf z}(x,t);\tau);\,\,
\Lambda = \sum_0^{2N}\lambda_j;
\label{AI}
\ee
$$\tau = (\tau_{ij}) = (\oint_{B_i}d\omega_j);\,\,\tau^{*}_{ij} = -\tau_{ij};
\,\,\Gamma = -2\sum_1^N\oint_{A
_j}\lambda d\omega_j$$
and ${\bf z}(x,t) = -2i[{\bf c}^N(x-x_0) + 2(\Lambda{\bf c}^N +
2{\bf c}^{N-1})t] + {\bf d}$ where $({\bf c}^N)_i = c_{iN}$ arises 
from the representation $d\omega_i = 
(\sum_1^N c_{ij}\lambda^{j-1})[d\lambda/R(\lambda)]$ (${\bf d}$ is a 
constant whose value is not important here).  Then the wave number
and frequency vectors can be defined via $\vec{\kappa} = -4i\pi
\tau^{-1}{\bf c}^N$ and $\vec{w} = -8i\pi\tau^{-1}[\Lambda
{\bf c}^N + 2{\bf c}^{N-1}]$ with $\theta_j(x,t) = \kappa_j x
+ w_j t + \theta_j^0$ (where the $\theta_j^0$ represent initial phases).
\\[3mm]\indent
To model the modulated wave now one writes
now $q = q_N(\theta_1,\cdots,\theta_N;\vec{\lambda})$ where $\lambda_j
\sim\lambda_j(X,T)$ and $\vec{\lambda}\sim(\lambda_j)$.
Then consider the first $2N+1$ polynomial
conservation laws arising from (\ref{AF}) - (\ref{AG})
and ${\bf C}$ for example (cf. below for KP)
and write these as $\partial_t{\cal T}_j(q) 
+ \partial_x{\cal X}_j(q) = 0$ (explicit formulas are given in 
\cite{fa} roughly as follows).
We note that the adjoint linear KdV equation (governing
the evolution of conserved densities) 
is $\partial_t\gamma_j + \partial^3_x\gamma_j - 6q\partial_x
\gamma_j = 0\,\,(\gamma_j\sim\nabla H_j)$ and (\ref{AG}) has the form
$\partial\gamma_{j+1} = (-(1/2)\partial^3 + q\partial +\partial q)\gamma_j$.
One then rewrites this to show that $6q\partial_x\gamma_j = \partial_x
[6\gamma_{j+1}-6q\gamma_j+3\partial^2\gamma_j]$ so that the adjoint
equation becomes
\be
\partial_t\gamma_j +\partial[-2\partial^2\gamma_j + 6q\gamma_j
-6\gamma_{j+1}]=0
\label{NF}
\ee
which leads to (\ref{AJ}) and (\ref{AN}) below (after simplification
of (\ref{NF})).  
Then for
the averaging step, write $\partial_t=\epsilon\partial_T$, etc.,
and average over the fast variable $x$ to obtain
\be
\partial_T<{\cal T}_j(q_N)> + \partial_X<{\cal X}_j(q_N)> = 0
\label{AJ}
\ee
((\ref{AJ}) makes the first order term in $\epsilon$ vanish).
The procedure involves averages
\be
<{\cal T}_j(q_N)> = lim_{L\to\infty}\frac{1}{2L}
\int_{-L}^L{\cal T}_j(q_N)dx
\label{AK}
\ee
for example (with a similar expression for $<{\cal X}_j(q_N)>$) and an
argument based on ergodicity is used.  Thus if the wave numbers
$\kappa_j$ are incommensurate the trajectory $\{q_N(x,t);\,\,x\in (-\infty,
\infty)\}$ will densely cover the torus ${\cal M}$.  Hence we can replace
$x$ averages with 
\be
<{\cal T}_j(q_N)> = \frac{1}{(2\pi)^N}\int_0^{2\pi}\cdots\int_0^{2\pi}
{\cal T}_j(q_N(\vec{\theta}))\prod_1^Nd\theta_j
\label{AL}
\ee
For computational purposes one can change the $\theta$ integrals to
$\mu$ integrals and obtain simpler calculations. By this procedure
one obtains a system of $2N+1$ first order partial differential equations
for the $2N+1$ points $\lambda_j(X,T)$, or equivalently for the
physical characteristics $(\vec{\kappa}(X,T),\vec{w}(X,T))$ (plus
$<q_N>$).
One can think of freezing the slow variables in the averaging and it is
$\underline{assumed}$ that (\ref{AJ}) is the correct first order description of
the modulated wave. 
\\[3mm]\indent
The above argument may or may not
have sounded convincing but it was in any case rather loose.
Let us be more precise following \cite{fa}.  One looks at the KdV
Hamiltonians beginning with $H = H(q) = lim_{L\to\infty}(1/2L)\int^L_
{-L}(q^2 + (1/2)q_x^2)dx$ (this form is appropriate for quasi-periodic
situations).  Then $q_t = \{q,H\}$ where $\{f,g\} = lim_{L\to\infty}
(1/2L)\int^L_{-L}(\delta f/\delta q)\partial_x(\delta g/\delta q)dx$ 
(averaged Gardner bracket).  The other Hamiltonians are found via
\be
\partial\frac{\delta H_{m+1}}{\delta q} = (q\partial + \partial q -
\frac{1}{2}\partial^3)\frac{\delta H_m}{\delta q}\,\,(m\geq 0);
\,\,\frac{\delta H_0}{\delta q} = 1
\label{AM}
\ee
where $\gamma_j\sim\nabla H_j\sim(\delta H_j/\delta q)$
(cf. here \cite{ca,fa}).  It is a general situation in the study of 
symmetries and conserved gradients (cf. \cite{cd}) that symmetries
will satisfy the linearized KdV equation $(\partial_t -6\partial_x q +
\partial^3_x)Q = 0$ and conserved gradients will satisfy the adjoint
linearized KdV equation $(\partial_t - 6q\partial_x + \partial^3_x)Q^{\dagger}
=0$; the important thing to notice here is that one is linearizing about
a solution $q$ of KdV.  Thus in our averaging processes the function
$q$, presumed known, is inserted in the integrals.  This leads then to
\be
{\cal T}_j(q) = \frac{\delta H_j}{\delta q};\,\,{\cal X}_j(q) =
-2\partial^2_x\frac{\delta H_j}{\delta q} - 6\frac{\delta H_{j+1}}{\delta q}
+ 6q\frac{\delta H_j}{\delta q}
\label{AN}
\ee
with (\ref{AJ}) holding, where $<\partial^2\phi>=0$ implies
\be
<{\cal X}_j> = lim_{L\to\infty}\frac{1}{2L}\int^L_{-L}(-6\frac
{\delta H_{j+1}}{\delta q_N} + 6q_N\frac{\delta H_j}{\delta q_N})dx
\label{AO}
\ee
In \cite{fa} the integrals are then simplified in terms of $\mu$ integrals
and expressed in terms of abelian differentials.  This is a beautiful
and important procedure linking the averaging process to the Riemann 
surface and is summarized in
\cite{mc} as follows.
One defines differentials 
\be
\hat{\Omega}_1 = -\frac{1}{2}[\lambda^N - \sum_1^Nc_j\lambda^{j-1}]\frac
{d\lambda}{R(\lambda)}
\label{AP}
\ee
$$
\hat{\Omega}_2 = [-\frac{1}{2}\lambda^{N+1} + \frac{1}{4}(\sum\lambda_j)
\lambda^N + \sum_1^N E_j\lambda^{j-1}]\frac{d\lambda}{R(\lambda)}
$$
where the $c_j,\,\,E_j$ are determined via 
$\oint_{b_i}\hat{\Omega}_1 = 0=\oint_{b_i}\hat{\Omega}_2\,\,(i = 1,
2,\cdots,N)$.  Then it can be shown that
\be
<\Psi>\sim
<{\cal T}>\sim\sum_0^{\infty}\frac{<{\cal T}_j>}{(2\mu)^j};\,\,
<{\cal X}>\sim\sum_0^{\infty}\frac{<{\cal X}_j>}{(2\mu)^j}
\label{NZ}
\ee
with
$\hat{\Omega}_1\sim<{\cal T}>(d\xi/\xi^2)$ and $<{\cal X}>(d\xi/\xi^2)\sim
12[(d\xi/\xi^4)-\hat{\Omega}_2]$ where $\mu=\xi^{-2}\to\infty\,\,(\mu\sim
(1/\sqrt{\xi})^{1/2}$) so $d\mu = -2\xi^{-3}d\xi\Rightarrow(d\xi/\xi^2)
\sim -(\xi/2)d\mu\sim-(d\mu/2\sqrt{\mu})$.  Since $\hat{\Omega}_1 =
O(\mu^N/\mu^{N+(1/2)})d\mu = O(\mu^{-(1/2)}d\mu,\,\,
\hat{\Omega}_2 =
O(\mu^{1/2})d\mu$ (with lead term $-(1/2)$) we obtain 
$<\Psi>\sim <{\cal T}>=O(1)$ and $<{\cal X}> = O(1)$.
Thus
(\ref{AF}), (\ref{AG}), ({\bf C}) generate all conservation laws
simultaneously with $<{\cal T}_j>$ (resp. $<{\cal X}_j>$) giving rise to
$\hat{\Omega}_1$ (resp. $\hat{\Omega}_2$).
It is then proved that all of the modulational
equations are determined via the equation 
\be
\partial_T\hat{\Omega}_1 = 12\partial_X\hat{\Omega}_2
\label{AR}
\ee
where the Riemann surface is thought of as depending on $X,T$ through
the points $\lambda_j(X,T)$.
In particular if the first $2N+1$ averaged conservation laws are satisfied
then so are all higher averaged conservation laws.  These equations
can also be written directly in terms of the $\lambda_j$ as Riemann
invariants via $\partial_T\lambda_j = S_j\partial_X\lambda_j$ for
$j = 0,1,\cdots,2N$ where $S_j$ is a computable characteristic speed
(cf. (\ref{FRR})-(\ref{FSS})).
Thus we have displayed the prototypical model for the Whitham or
modulational equations.
\\[3mm]\indent
Another way of looking at some of this goes as follows.
We will consider surfaces defined via $R(\Lambda)=
\prod_1^{2g+1}(\Lambda-\Lambda_i)$.
For convenience take the branch points $\Lambda_i$ real with $\Lambda_1
<\cdots<\Lambda_{2g+1}<\infty$.  This corresponds to spectral bands
$[\Lambda_1,\Lambda_2],\cdots,[\Lambda_{2g+1},\infty)$ and gaps 
$(\Lambda_2,\Lambda_3),\cdots,(\Lambda_{2g},\Lambda_{2g+1})$ with the
$A_i$ cycles around the gaps (i.e. $a_i\sim (\Lambda_{2i},\Lambda_{2i+1}),
\,\,i=1,\cdots,g$).
The notation is equivalent to what preceeds with a shift of index.
For this kind of situation one usually defines
the period matrix via $iB_{jk}=\oint_{B_k}\omega_j$ and sets
\be
\oint_{A_k}d\omega_j = 2\pi\delta_{jk}\,\,(j,k=1,\cdots,g);\,\,
d\omega_j = \sum_1^g\frac{c_{jq}\Lambda^{q-1}d\Lambda}{\sqrt{R(\Lambda)}}
\label{FHH}
\ee
(cf. \cite{bb,ca,cn}).
The $B_i$ cycles can be drawn from a common vertex ($P_0$ say) passing
through the gaps $(\Lambda_{2i},\Lambda_{2i+1})$.  One chooses now e.g.
$p=\int dp$ and $\Omega = \int d\Omega$ in the form
\be
p(\Lambda) = \int dp(\Lambda) = \int \frac{{\cal P}(\Lambda)d\Lambda}
{2\sqrt{R(\Lambda)}};\,\,{\cal P} = \Lambda^g+\sum_1^ga_j\Lambda^{g-j};
\label{FFF}
\ee
$$\Omega(\Lambda) = \int d\Omega(\Lambda) = \int\frac{6\Lambda^{g+1} 
+{\cal O}(\Lambda)}{\sqrt{R(\Lambda}}d\Lambda;\,\,{\cal O} = 
\sum_0^gb_j\Lambda^{g-j};\,\,b_0 = -3\sum_1^{2g+1}\Lambda_i$$
with normalizations
$\int^{\Lambda_{2i+1}}_{\Lambda_{2i}}dp(\Lambda) = \int^{\Lambda_{2i+1}}_
{\Lambda_{2i}}d\Omega(\Lambda) = 0;\,\,i=1,\cdots,g$.
We note that one is thinking here of (*) $\psi=exp[ipx+i\Omega t]\,\,\cdot$
theta functions (cf. (\ref{psi}) with e.g. 
$p(\Lambda) = -i\overline{(log\psi)_x};\,\,\Omega(\Lambda) = 
-i\overline{(log\psi)_t}$.
Recall that the notation $<\,,\,>_x$ simply
means $x$-averaging (or ergodic averaging)
and $\overline{(\log\psi)_x}\sim <(log\psi)_x>_x\not= 0$ here since e.g.
$(log\psi)_x$ is not bounded.
Observe that
(*) applies to any finite zone quasi-periodic situation.  
The KdV equation here arises from
$L\psi=\Lambda\psi,\,\,L=-\partial^2+q,\,\,\partial_t\psi=A\psi,\,\,
A=4\partial^3-6q\partial-3q_x$ and there is no need to put this
in a more canonical form since this material is only illustrative.
\\[3mm]\indent
In this context one has also the Kruskal integrals
$I_0,\cdots,I_{2g}$ which arise via a generating function
\be
p(\Lambda) = -i<(log\psi)_x>_x = \sqrt{\Lambda} + \sum_0^{\infty}
\frac{I_s}{(2\sqrt{\Lambda})^{2s+1}}
\label{FLL}
\ee
where $I_s = <P_s>_x = \overline{P_s}
\,\,(s = 0,1,\cdots$) with $-i(log\psi)_x =
\sqrt{\Lambda} + \sum_0^{\infty}[P_s/(2\sqrt{\Lambda})^{2s+1}]$.
Similarly 
\be
-i\overline{(log\psi)_t} = -i\frac{A\psi}{\psi} = 4(\sqrt{\Lambda})^3 +
\sum_0^{\infty}\frac{\Omega_s}{(2\sqrt{\Lambda})^{2s+1}}
\label{FMM}
\ee
and one knows $\partial_tP_s = \partial_x\Omega_s$ since 
$(\spadesuit)\,\, [(log\psi)_x]_t =
[(log\psi)_t]_x$.  The expansions are standard (cf. \cite{ch,ci,cl,cn}).
Now consider a ``weakly deformed" soliton lattice of the form 
\be
\theta(\tau|B)=\sum_{-\infty<n_1<\cdots<n_m<\infty}exp(-\frac{1}{2}
\sum_{j,k}B_{jk}n_jn_k + i\sum_jn_j\tau_j)
\label{FII}
\ee
with the $\Lambda_i\,\,(i=1,\cdots,2g+1)$ (or equivalently the parameters
$u^i = I_i;\,\,i=0,\cdots,2g$) slowly varying functions of $x,t$ (e.g.
$u^i = u^i(X,T),\,\,X = \epsilon x,\,T=\epsilon t,\,\,i = 0,1,\cdots,2m$).
Now one wants to obtain a version of (\ref{AR}) directly via
$(\spadesuit)$.  Thus insert the slow variables in $(\spadesuit)$ and average,
using $\epsilon\partial_X$ or $\epsilon\partial_T$ in the external derivatives,
to obtain
\be
\partial_T\overline{(log\psi)_x} = \partial_X\overline{(log\psi)_t}
\label{FOO}
\ee
or $\partial_Tp(\Lambda) = \partial_XE(\Lambda)$.  Then from (\ref{FFF})
differentiating in $\Lambda$ one gets
\be
\partial_Tdp = \partial_Xd\Omega
\label{FPP}
\ee
Now (recall $\partial_tP_s = \partial_x
\Omega_s$) expanding (\ref{FPP}) in powers of $(\sqrt{\Lambda})^{-1}$ one
obtains the slow modulation equations in the form
\be
\partial_T u^s = \partial_X\bar{\Omega}_s\,\;\,(s=0,\cdots,2g)
\label{FQQ}
\ee
where $\overline{\Omega_s}$ is a function of the $u^i\,\,(0\leq i\leq 2g)$.
This leads to equations
\be
\partial_T\Lambda_i = v_i(\Lambda_1,\cdots,\Lambda_{2g+1})\partial_X
\Lambda_i\,\;\,(i=1,\cdots,2g+1)
\label{FRR}
\ee
for the branch points $\Lambda_k$ as Riemann invariants.  The characteristic
``velocities" have the form
\be
v_i = \left.\frac{d\Omega}{dp}\right|_{\Lambda=\Lambda_i} =
2\frac{6\Lambda_i^{g+1} +
\Omega(\Lambda_i)}{p(\Lambda_i)}\,\;\,(1\leq i\leq 2g+1)
\label{FSS}
\ee
To see this simply multiply (\ref{FPP}) by $(\Lambda-\Lambda_i)^{3/2}$ and
pass to limits as $\Lambda\to\Lambda_i$.  

\subsection{Extension to KP}
Given some knowledge of symmetries as sketched in \cite{cd} for example
one is tempted to rush now to an immediate attempt at generalizing the
preceeding results to finite zone KP via the following facts.  The KP
flows can be written as $\partial_nu = K_n(u)$ where the $K_n$ are 
symmetries satisfying (in the notation of \cite{cd})
the linearized KP equation $\partial_3\beta = (1/4)\partial^3\beta
+ 3\partial(u\beta) + (3/4)\partial^{-1}\partial_2^2\beta = K'[\beta]$.
The conserved densities or gradients 
$\gamma$ satisfy the adjoint linearized KP
equation $\partial_3\gamma = (1/4)\partial^3\gamma + 3u\partial\gamma
+ (3/4)\partial^{-1}\partial_2^2\gamma$.  Then, replacing the square
eigenfunctions by $\psi\psi^{*}$ one has e.g. $\psi\psi^{*} = \sum_0^
{\infty}s_n\lambda^{-n}$ where $s_n\sim$ a $\gamma_n$.  
Further $\partial_nu
= K_{n+1} = \partial s_{n+1} = \partial Res\,L^n = \partial\nabla\hat
{I}_{n+1}$ where $\nabla f\sim\delta f/\delta u$
($Res\,L^n = nH^1_{n-1}$ is generally used in the multipotential
theory).  We are working here in a single potential theory where all
potentials $u_i$ in $L = \partial + \sum_1^{\infty}u_{i+1}\partial^{-i}$
are expressed in terms of $u_2=u$ via operators with $\partial$ and
$\partial^{-1}$.  One uses here the Poisson bracket $\{f,g\} =
\int\int(\delta f/\delta u)\partial(\delta g/\delta u)dxdy$ (Gardner
bracket).   Let us retrace
the argument from \cite{fa} and see what applies for KP.  Thus 
one has $s_{n+1}\sim\gamma_{n+1}\sim\nabla\hat{I}_{n+1}$
as conserved gradients satisfying the adjoint linear KP equation
(**) $\partial_t\gamma = (1/4)\partial^3\gamma + 3u\partial\gamma
+(3/4)\partial^{-1}\partial^2_y\gamma$.  The nonlocal term $\partial^{-1}$ here
could conceivably change some of the analysis.
We need first a substitute for (\ref{AG}) or
else a direct way of rewriting the adjoint equation as 
perhaps $\partial_t[A]+\partial_x[B]
+\partial_y[C] =0$.  To get such a formula one thinks of
differentiating the adjoint linearized KP equation
to get
\be
\partial_t[\partial\gamma] +\partial[\frac{1}{4}\partial^3\gamma +
3u\partial\gamma]+\partial_y[\frac{3}{4}\partial_y\gamma]=0
\label{NG}
\ee
This removes all of the nonlocal terms and one doesn't have to deal with
$\partial[(3/4)\partial^{-1}\partial_y^2\gamma]$ for example;
however it leads to a redundant situation.  
Let us try instead
\be
\partial_t[\gamma] + \partial[\frac{1}{4}\partial^2\gamma+3\partial^{-1}
u\partial\gamma] +\partial_y[\frac{3}{4}\partial_y\partial^{-1}\gamma]=0
\label{NJ}
\ee
One expects that
$<\partial^2\gamma>=0
=<\partial_y\partial^{-1}\gamma>$ and we take
$\hat{{\cal L}} = 3\partial^{-1}(u_0\partial
\gamma)$ so that,
averaging as in KdV, one obtains
\be
\partial_T<\gamma> + \partial_X<\hat{{\cal L}}> 
=0
\label{NS}
\ee
which conceivably might be useful.  This is discussed later.

\section{AVERAGING WITH $\psi^*\psi$}
\renewcommand{\theequation}{3.\arabic{equation}}\setcounter{equation}{0}

Let us look now at \cite{ka} but in the spirit of \cite{fb}.
We will expand upon this with some modifications in order to obtain  
a visibly rigorous procedure (cf. also \cite{co}).  Thus
consider KP in the form (\ref{AC}): $3\sigma^2u_{yy} + \partial_x(4u_t -
6uu_x + u_{xxx}) = 0$ via compatibility $[\partial_y -L,\partial_t - A]
=0$ where $L = \sigma^{-1}(\partial^2-u)$ and $A = \partial^3 - (3/2)
u\partial + w\,\, (\sigma^2 = 1$ is used in \cite{fb} which we follow for
convenience but the procedure should work in general with minor
modifications - note $\partial$ means $\partial_x$ and $u\to -u$ in the
development of (\ref{psi})).  We have then
$(\partial_y - L)\psi = 0$ with $(\partial_t -A)\psi = 0$ and for the 
adjoint or dual wave function 
$\psi^{*}$ one writes in \cite{ka} $\psi^{*}L = -\partial_y\psi^{*}$
with $\psi^{*}A = \partial_t\psi^{*}$ where $\psi^{*}(f\partial^j)\equiv
(-\partial)^j(\psi^{*}f)$.  We modify the formulas used in \cite{ka}
(and \cite{fb}) in taking
\be
\psi = e^{px+Ey+\Omega t}\cdot\phi(Ux+Vy
+Wt,P);\,\,
\psi^* = e^{-px-Ey-\Omega t}\cdot\phi^*
(-Ux-Vy-Wt,P)
\label{YE}
\ee
where $p=p(P),\,\,E = E(P),\,\,\Omega =
\Omega(P),$ etc. (cf. (\ref{YDD}) - (\ref{YEE})), which isolate
the quantities needed in averaging. 
The arguments to follow are essentially the same for this
choice of notation, or that in \cite{fb} or \cite{ka}.
Now one sees immediately that 
\be
(\psi^{*}L)\psi = \psi^{*}L\psi + \partial_x(\psi^{*}L^1\psi) +
\partial^2_x(\psi^{*}L^2\psi) + \cdots
\label{AS}
\ee
where e.g. $L^r = ((-1)^r/r!)(d^r L/d(\partial)^r)$.  In particular
\be
\psi^*_{xx}\psi =\psi^*\psi_{xx} -2\partial(\psi^*\psi_x) +
\partial^2(\psi^*\psi);
\label{YG}
\ee
$$-\psi^*_{xxx}\psi = \psi^*\psi_{xxx} -\partial^3(\psi^*\psi)
+3[\partial^2(\psi^*\psi_x) - \partial(\psi^*\psi_{xx})]
$$
This means ($L^*=L,\,\,\psi^*A\sim A^*\psi^* =-\partial^3\psi^* 
+(3/2)\partial(\psi^*u)+w\psi^*$)
\be
(\psi^*L)\psi = [(\partial^2-u)\psi^*]\psi = \psi^*_{xx}\psi -u\psi^*\psi
=\psi^*L\psi -2\partial(\psi^*\psi_x) +\partial^2(\psi^*\psi)
\label{YI}
\ee
which implies $L^1 = -2\partial$ and $L^2 = 1$.  Next
\be
(\psi^*A)\psi = -\psi^*_{xxx}\psi +\frac{3}{2}(\psi^*_xu+\psi^*u_x)\psi
+w\psi^*\psi = \psi^*(A\psi) +
\label{YJ}
\ee
$$+ \partial[\psi^*(\frac{3}{2}u-3\partial^2)\psi] + 3\partial^2(\psi^*\psi_x)
-\partial^3(\psi^*\psi)$$
so that $A^1 = -3\partial^2+(3/2)u,\,\,A^2 = 3\partial,$ and $A^3 =
-1$.
We think of a general Riemann surface $\Sigma_g$.  Here one picks
holomorphic differentials 
$d\omega_k$ as before and quasi-momenta, quasi-energies, etc. via
$dp\sim d\Omega_1,\,\,dE\sim d\Omega_2,\,\,d\Omega\sim d\Omega_3,\cdots$
where $\lambda\sim k,\,\,p = \int^P_{P_0}d\Omega_1,$ etc.
Normalize
the $d\Omega_k$ so that $\Re\int_{A_i}d\Omega_k = 0 = \Re\int_{B_j}
d\Omega_k$ (cf. remarks after (\ref{YEE})); then
$U,\,V,\,W,\cdots$ are real $2g$ period vectors.
and one has BA functions $\psi(x,y,t,P)$ as
in (\ref{YE}).  As before we look for approximations 
based on
$u_0(xU+yV+tW|I) = u_o(\theta_j,I_k)$.  For averaging, $\theta_j \sim
xU_j + yV_j + tW_j,\,\,1\leq j\leq 2g$,
with period $2\pi$ in the
$\theta_j$ seems natural (but note
$\theta_j,\,\theta_{g+j}\sim U_j,$ etc.).  
Then again by ergodicity
$<\phi>_x = lim_{L\to\infty}(1/2L)\int_{-L}^L\phi dx$ becomes
$<\phi> = (1/(2\pi)^{2g}\int\cdots\int\phi d^{2g}\theta$ and one
notes that $<\partial_x\phi> = 0$ automatically
for $\phi$ bounded.  In \cite{fb} one thinks
of $\phi(xU+\cdots)$ with $\phi_x = \sum U_i(\partial\phi/\partial\theta_i)$
and $\int\cdots\int(\partial
\phi/\partial\theta_i)d^{2g}\theta = 0$. 
\\[3mm]\indent
Now for averaging we think of $u_0\sim u_0(\frac{1}{\epsilon}S|I)$ as
in (\ref{AE}) with $(\hat{S},I)\sim (\hat{S},I)(X,Y,T),
\newline
\partial_X \hat{S} = U,\,\,\partial_Y
\hat{S} = V,$ and $\partial_T\hat{S} = W$.  We think of expanding
about $u_0$ with $\partial_x\to\partial_x + \epsilon\partial_X$.
This step will cover both $x$ and $X$ dependence for subsequent averaging.  
Then look at the compatibility condition $(\clubsuit):\,\,\partial_tL
-\partial_y A + [L,A] = 0$.  As before we will want the term of first
order in $\epsilon$ upon writing e.g. $L = L_0 + \epsilon L_1 + \cdots$
and $A = A_0 + \epsilon A_1 + \cdots$ where slow variables appear only
in the $L_0,\,A_0$ terms.
As indicated in \cite{ka} the term
$[L,A]\to \{L,A\}$ where $\{L,A\}$ arises upon replacing $\partial_x$
be $\partial_x + \epsilon\partial_X$ in all the differential expressions
and taking the terms of first order in $\epsilon$.  However the formula
in \cite{ka} is somewhat unclear so we compute some factors explicitly.
In fact,
according to \cite{fb}, one can 
write now, to make the coefficient of $\epsilon$ vanish
\be
\partial_t L_1 - \partial_y A_1 + [L_0,A_1] + [L_1,A_0] + F = 0;\,\,
F=\partial_TL - \partial_YA - (L^1\partial_XA - A^1\partial_XL)
\label{AT}
\ee
Thus $F$ is the first order term involving derivatives in the slow
variables (note the $L^1\partial_XA$ and $A^1\partial_XL$ terms have
opposite signs from \cite{fb,ka} but this seems to be correct).
To clarify this let us write $(\hat{\partial}\sim\partial/\partial X)\,\,
L_{\epsilon} = (\partial +\epsilon
\hat{\partial})^2 - (u_0+\epsilon u_1+\cdots) = \partial^2-u_0 +\epsilon
(2\partial\hat{\partial}-u_1) +O(\epsilon^2)$ and $A_{\epsilon} = 
(\partial+\epsilon\hat{\partial})^3 -(3/2)(u_0+\epsilon u_1+\cdots)\cdot
(\partial+\epsilon\hat{\partial}) + w_0+\epsilon w_1+\cdots = \partial^3
-(3/2)u_0\partial+w_0+\epsilon(3\partial^2\hat{\partial} -
(3/2)u_1\partial -(3/2)u_0\hat{\partial} + w_1) + O(\epsilon^2)$.  Write
then $A_{\epsilon} = A_0+\hat{A}_1\epsilon + O(\epsilon^2)$ and $L_{\epsilon}
= L_0 +\hat{L}_1\epsilon +O(\epsilon^2)$ with $\hat{A}_1 = 3\partial^2
\hat{\partial}-(3/2)u_1\partial-(3/2)u_0\hat{\partial} + w_1$ and
$\hat{L}_1 = 2\partial\hat{\partial}-u_1$.  Note that $L^1 = -2\partial\not=
\hat{L}_1$ and $A^1 = -3\partial^2+(3/2)u_0\not=\hat{A}_1$ but we can write
\be
\hat{A}_1 = -A^1\hat{\partial}-(3/2)u_1\partial+w_1 = -A^1\hat{\partial}
+A_1;
\label{YL}
\ee
$$\hat{L}_1 = -L^1\hat{\partial}-u_1 = -L^1\hat{\partial}+L_1$$
with $L_1=-u_1$ and $A_1=
-(3/2)u_1\partial+w_1$ as in \cite{fb}.  Then $(\clubsuit)$ becomes
\be
\partial_tL_0 -\partial_yA_0 +[L_0,A_0] +\epsilon\{\partial_tL_1
-\partial_yA_1 +\partial_TL-\partial_YA +[L_0,\hat{A}_1] +[\hat{L}_1,A_0]\}
+O(\epsilon^2)
\label{YK}
\ee
Now the $\partial_t
L_0-\partial_yA_0+[L_0,A_0]$ term vanishes and we note that
\be
[L_0,\hat{A}_1]+[\hat{L}_1,A_0] = [L_0,-A^1\hat{\partial}] +
[-L^1\hat{\partial},A_0] + [L_0,A_1]+[L_1,A_0] =
\label{YM}
\ee
$$= [L_0,A_1]+[L_1,A_0] +A^1\hat{\partial}L_0-L^1\hat{\partial}A_0
-L_0A^1\hat{\partial} +A_0L^1\hat{\partial}$$
Then one notes that $\partial_XL_0\sim\partial_XL$ and $\partial_XA_0
\sim\partial_XA\,\,(\partial_X\sim\hat{\partial}$) so dropping the 
terms in (\ref{YM}) with an inoperative $\hat{\partial}$ on the right
we obtain (\ref{AT}).
Next one writes, using (\ref{AS})
\be 
\partial_t(\psi^*L_1\psi) -\partial_y(\psi^*A_1\psi) = \psi^*\{L_{1t}-A_{1y} 
+[L_1,A] +[L,A_1]\}\psi = 
\label{AU}
\ee
$$ = \psi^*(\partial_tL_1-\partial_yA_1 + [L_0,A_1] + [L_1,A_0])\psi +
\partial_x(\cdots)$$
and via ergodicity in $x,y$, or $t$ flows, averaging of derivatives in
$x,y$, or $t$ gives zero, so from (\ref{AT}) and (\ref{AU}) we obtain
the Whitham equations in the form $<\psi^*F\psi> = 0$
(this represents the first order term in $\epsilon$ - the slow variables
are present in $L_0,\,\,A_0,\,\,\psi,$ and $\psi^*$).  
In order to spell
this out in \cite{fb} one imagines $X,Y,T$ as a parameter $\xi$ and 
considers $L(\xi),\,\,A(\xi)$, etc. (in their perturbed form) with
\be
\psi(\xi) = e^{p(\xi)x+E(\xi)y+\Omega(\xi)t}\cdot
\phi(U(\xi)x + V(\xi)y 
+ W(\xi)t|I(\xi))
\label{AV}
\ee
where $\psi^* = exp(-px-Ey-\Omega t)\phi^*(-Ux-Vy-Wt|I)$ 
(no $\xi$ variation - i.e. assume $p,E,\Omega,U,V,W,I$ fixed).
We recall that
one expects $\lambda_k = \lambda_k(X,Y,T)$ etc. so the Riemann surface
varies with $\xi$.  Also recall that
$x,y,t$ and $X,Y,T$ can be considered as independent variables.  Now as
above, using (\ref{AS}), we can write
\be
\partial_t(\psi^*\psi(\xi)) = \psi^*(A(\xi)-A)\psi(\xi) - \partial_x
(\psi^*A^1\psi(\xi)) + \partial^2_x(\cdots)
\label{AW}
\ee
Note also from (\ref{AV}) for $P$ fixed
($\theta\sim xU+yV+yW$)
\be
\partial_{\xi}\psi^*\psi(\xi)|_{\xi=0}= (\dot{p}x+\dot{E}y+\dot{\Omega}t)
\psi^*\psi + 
\label{AX}
\ee
$$ + (\dot{U}x+\dot{V}y +\dot{W}t)\cdot\psi^*\partial_{\theta}\psi +
\dot{I}\cdot\psi^*\partial_I\psi$$
where $\dot{f}\sim\partial f/\partial\xi$.  In \cite{fb} one assumes
that it is also permitted to vary $\xi$ and hold
e.g. the $I_k$ constant while allowing say the $P$ to vary.
Now differentiate the left side $\partial_t
(\psi^*\psi(\xi))$ of (\ref{AW}) in $\xi$ and use (\ref{AX}) to obtain
\be
\partial_{\xi}[\partial_t(\psi^*\psi(\xi))]|_{\xi=0} = \dot{\Omega}
\psi^*\psi + \dot{W}\cdot\psi^*\psi_{\theta} +
\label{AY}
\ee
$$ + \{(\dot{p}x+\dot{E}y+\dot{\Omega}t)\partial_t(\psi^*\psi) +
(\dot{U}x+\dot{V}y+\dot{W}t)\cdot\partial_t(\phi^*\phi_{\theta}) +
\dot{I}\cdot\partial_t(\phi^*\phi_I)\}$$
Fixing $x,y,t$ and averaging (\ref{AY}) in the $\theta$ variables yields
\be
<\partial_{\xi}[\partial_t(\psi^*\psi(\xi))]|_{\xi=0}> =
\dot{\Omega}<\psi^*\psi> + \dot{W}<\psi^*\psi_{\theta}>
\label{AZ}
\ee
Next one differentiates the right side of (\ref{AW}) in $\xi$, averages
in the $\theta$ variables for $x,y,t$ fixed,
and equates to (\ref{AZ}), to obtain
\be
(\ref{AZ}) = <\psi^*\dot{A}\psi> - <\partial_{\xi}[\partial_x(\psi^*
A^1\psi(\xi))]|_{\xi=0}> = 
\label{BA}
\ee
$$<\psi^*\dot{A}\psi> - \dot{p}<\psi^*A^1\psi>
-\dot{U}\cdot <\psi^*A^1\psi_{\theta}>$$
We note here as in (\ref{AX}) - (\ref{AY}) (one can disregard $A^1$ action 
on the isolated $x$ term, 
and assumes $I$ fixed, since only more terms $\partial_x(\cdots)$ 
arise whose average vanishes)
\be
\partial_x\partial_{\xi}(\psi^*A^1\psi)|_{\xi=0} = \partial_x[\psi^*
A^1(\dot{p}x+\dot{E}y+\dot{\Omega}t)\psi +
\label{BB}
\ee
$$ + \psi^*A^1(\dot{U}x+\dot{V}y+\dot{W}t)\cdot\psi_{\theta}] = 
\dot{p}\psi^*A^1\psi + \dot{U}\cdot\psi^*A^1\psi_{\theta} +$$
$$+ (\dot{p}x+\dot{E}y+\dot{\Omega}y)\partial_x(\psi^*A^1\psi) +
(\dot{U}x+\dot{V}y+\dot{W}t)\cdot\partial_x(\psi^*A^1\psi_{\theta})$$
and (\ref{BA}) follows.  Rewriting (\ref{BA}) with $\xi\sim Y$ we obtain
(note $\psi^*\psi_{\theta} = \phi^*\phi_{\theta}$)
\be
-<\psi^*\partial_YA\psi> = -\partial_Y\Omega<\psi^*\psi> -\partial_YW
<\phi^*\phi_{\theta}> -
\label{BC}
\ee
$$ -\partial_Yp<\psi^*A^1\psi> -\partial_YU\cdot<\psi^*A^1
\psi_{\theta}>$$
Similarly one uses (cf. (\ref{AW}))
\be
\partial_y(\psi^*\psi(\xi)) = \psi^*(L(\xi)-L)\psi(\xi) - \partial_x
(\psi^*L^1\psi(\xi)) + \partial_x^2(\cdots)
\label{BD}
\ee
with $\xi\sim T$ to get
\be
<\psi^*\partial_TL\psi> = \partial_TE<\psi^*\psi> + \partial_TV\cdot
<\phi^*\phi_{\theta}> +
\label{BE}
\ee
$$ + \partial_Tp<\psi^*L^1\psi> +\partial_TU\cdot<\psi^*L^1\psi_{\theta}>$$
Finally note that
\be
\partial_y(\psi^*A^1\psi(\xi)) -\partial_t(\psi^*L^1\psi(\xi)) =
\label{BF}
\ee
$$ = \psi^*[L^1(A(\xi)-A) - A^1(L(\xi)-L)]\psi(\xi) + \partial_x(\cdots)$$
Using this with $\xi\sim X$ one gets then as above 
$$<\psi^*(L^1\partial_XA - A^1\partial_XL)\psi> = \partial_X\Omega
<\psi^*L^1\psi> - \partial_XE<\psi^*A^1\psi> +$$
\be
+ \partial_XW\cdot<\psi^*L^1\psi_{\theta}> - \partial_XV\cdot
<\psi^*A^1\psi_{\theta}>
\label{BG}
\ee
This formula (\ref{BG}) will be of importance later in
deciphering the notation of \cite{kt}.
We recall that $\partial_X\hat{S} = U,\,\,
\partial_Y\hat{S} = V,$ and $\partial_T\hat{S}
= W$ so there are compatibility relations
\be
\partial_YU = \partial_XV;\,\,\partial_TU = \partial_XW;\,\,\partial_TV
= \partial_YW
\label{BH}
\ee
Now add up (\ref{BC}) and (\ref{BE}), and subtract (\ref{BG}) to get
\be
<\psi^*[\partial_TL - \partial_YA - L^1\partial_XA + A^1\partial_XL]\psi>
= <\psi^*F\psi> = 0 =
\label{BI}
\ee
$$ = (\partial_TE-\partial_Y\Omega)<\psi^*\psi> + (\partial_Tp -
\partial_X\Omega)<\psi^*L^1\psi> + (\partial_XE - \partial_Yp)
<\psi^*A^1\psi>$$
We observe that if one lets the point $P$ on $\Sigma$ vary with $\xi$,
while holding $\theta_k$ and $I_j$ fixed, then
\be
\partial_{\xi}(\psi^*\psi(\xi))|_{\xi=0} = (xdp + ydE + td\Omega)\phi^*
\phi
\label{BJ}
\ee
which, together with (\ref{AW}) and (\ref{BD}), yields e.g.
\be
d\Omega<\phi^*\phi> = -dp<\psi^*A^1\psi>;\,\,dE<\phi^*\phi> = 
-dp<\psi^*L^1\psi>
\label{BK}
\ee
(cf. (\ref{BA}) - (\ref{BB})).  It follows then from (\ref{BI}) and
(\ref{BK}) that
\be
0 = (\Omega_Y-E_T)dp + (p_T-\Omega_X)dE + (E_X-p_Y)d\Omega
\label{BL}
\ee
Thus one arrives at a version of the Whitham equations in the form
\be
p_T = \Omega_X;\,\,p_Y = E_X;\,\,E_T = \Omega_Y
\label{BM}
\ee
where the last equation establishes compatibility of the first two via
$p_{TY} - p_{YT} = \Omega_{XY} - E_{XT} = \partial_X(\Omega_Y  - E_T) = 0$.
We feel that this derivation from \cite{fb}, based on \cite{ka}, 
is important since it
again exhibits again the role of square eigenfunctions (now in the form
$\psi^*\psi$) in dealing with averaging processes.  In view of the
geometrical nature of such square eigenfunctions (cf. \cite{cd} for
example) one might look for underlying geometrical objects related to
the results of averaging.  Another (new) direction 
involves the Cauchy kernels expressed via $\psi^*\psi$
and their dispersionless limits (cf. \cite{cq}). 
\\[3mm]\indent
Let us write now
($\Upsilon\sim px+Ey+\Omega t$ and $\Xi=xU+yV+tW$) with
$$
<\psi^*L^1\psi> = <\psi^*(-2\partial)\psi> = -2<e^{-\Upsilon}\phi^*
[pe^{\Upsilon}\phi +Ue^{\Upsilon}\phi_{\theta}]>=$$
\be
-2[p<\phi^*\phi> + U<\phi^*\phi_{\theta}>]
\label{PA}
\ee
Similarly from $A^1 = -3\partial^2+(3/2)u_0$ we have
\be
<\psi^*A^1\psi> = -3<\psi^*[p^2e^{\Upsilon}\phi+2pUe^{\Upsilon}\phi_
{\theta}+U^2e^{\Upsilon}\phi_{\theta\theta}]>+
\label{PB}
\ee
$$+\frac{3}{2}<\psi^*u_0\psi>=-3[p^2<\phi^*\phi>+2pU<\phi^*\phi_{\theta}>
+U^2<\phi^*\phi_{\theta\theta}>]+\frac{3}{2}<\phi^*u_0\phi>$$
We note also a natural comparison with (\ref{NZ}) for the formula
\be
<\psi\psi^*> = \sum_1^{\infty}\frac{<s_n>}{\lambda^n}
\label{PC}
\ee
We remark here that such a series is natural from asymptotic expansions
but when $\psi,\,\,\psi^*$ are written in terms of theta functions it
requires expansion of the theta functions in $1/\lambda$ (such
expansions are documented in \cite{dc}, p. 49 for example).  It is now
natural to ask whether one can express $dp,\,\,dE,\,\,d\Omega$ from
(\ref{BK}) in more detail.  Thus note first from (\ref{BK})
\be
dE<\phi^*\phi> = 2dp[p<\phi^*\phi> +U<\phi^*\phi_{\theta}>];\,\,
d\Omega<\phi^*\phi> =
\label{PD}
\ee
$$3dp[p^2<\phi^*\phi> + 2pU<\phi^*\phi_{\theta}> + U^2<\phi^*\phi_{\theta
\theta}>]-\frac{3}{2}dp<\phi^*u_0\phi>$$
which says e.g.
\be
dE = 2pdp + \frac{2U<\phi^*\phi_{\theta}>}{<\phi^*\phi>}dp;\,\,
d\Omega =
\label{PE}
\ee
$$= 3p^2dp +\left[6pU\frac{<\phi^*\phi_{\theta}>}{<\phi^*\phi>} +
3U^2\frac{<\phi^*\phi_{\theta\theta}>}{<\phi^*\phi>} -\frac{3}{2}
\frac{<\phi^*u_0\phi>}{<\phi^*\phi>}\right]dp$$
Given $p\sim\lambda+\cdots,\,\,E\sim\lambda^2+\cdots,\,\,
\Omega\sim\lambda^3+\cdots$ this is reasonable.  One would also like
to compare (\ref{NS}) and the equation $\partial_Tdp +\partial_Xd\Omega
+\partial_YdE=0$ (recall $\hat{\cal L}=(1/4)\partial^2\gamma+3\partial^{-1}
(u_0\partial\gamma)$ and $\hat{\cal G} = (3/4)\partial_y\partial^{-1}\gamma$).
One expects $dp = d\lambda+\cdots,\,\,dE = 2\lambda d\lambda+\cdots,$
and $d\Omega=3\lambda^2 d\lambda+\cdots$ so a formula like (\ref{PC})
has an order of magnitude compatibility with a possible 
asymptotic identification
$dp\sim<\gamma>d\lambda$.
Note here the analogy with (\ref{NZ}) etc. where $<\Psi>\sim <{\cal T}>$
with $\hat{\Omega}_1\sim -<{\cal T}>(d\mu/2\sqrt{\mu})$ (the $1/\sqrt{\mu}$
weight factor arises from KdV notation here, linked to a 
hyperelliptic Riemann surface); recall also $<{\cal X}>$ is
similarly connected to $\hat{\Omega}_2$.  Let us assume the ``ansatz"
$(\clubsuit\clubsuit)$:  For some choice of standard homology basis and
local coordinate, $dp\sim <\psi\psi^*>d\lambda$ for large $|\lambda|$. 
We have seen that a
variation on this is valid for KdV situations and 
via changes in homology and local
coordinate it is reasonable in general.  In fact it generally true via
\\[3mm]\indent {\bf LEMMA 3.1}.$\,\,$  The ansatz $(\clubsuit\clubsuit)$ is
valid for KP situations.
\\[2mm]\indent {\it Proof}:$\,\,$  We refer here to \cite{cd} for
background and notation (cf. also \cite{ci} for dispersionless
genus zero situations).  We write $\partial 
= L + \sum_1^{\infty}\sigma_j^1 L^{-j}$ which implies
$\partial\psi/\psi=\lambda + \sum_1^{\infty} 
\sigma_j^1\lambda^{-j}$. (this is motivated by the proof
of Lemma 2.1 in \cite{ba}).  Then 
$(log\psi)_x=\lambda +\sum_1^{\infty}\sigma_j^1\lambda^{-j}$ and
$\overline{(log\psi)_x} = p
= \lambda + \sum_1^{\infty} <\sigma_j^1>\lambda^{-j}$.  But $s_{n+1} =
-n\sigma_n^1 - \sum_1^{n-1}\partial_j\sigma_{n-j}^1$ and one 
obtains $<s_{n+1}> =
-n<\sigma_n^1> -\sum_1^{n-1}<\partial_j\sigma_{n-j}^1> = -n<\sigma_n^1>$. Thus,
$dp = d\lambda - \sum_1^{\infty}j<\sigma_j^1>\lambda^{-j-1}d\lambda = d\lambda
+\sum_1^{\infty}<s_{j+1}>\lambda^{-j-1}d\lambda=<\psi\psi^*>$ since
$\psi\psi^*=\sum_0^{\infty}s_n\lambda^{-n}$, and the
normalization is built into this construction.  $\,\,\,$
{\bf QED}
\\[3mm]\indent
Similarly from
(\ref{BK}) we obtain for large $\lambda$
\be
d\Omega\sim -<\psi^*A^1\psi>d\lambda;\,\,dE\sim -<\psi^*L^1\psi>d\lambda
\label{PF}
\ee
(note this requires $<\psi^*L^1\psi> = O(2\lambda)$ which by (\ref{PA})
is correct while $<\psi^*A^1\psi>$ should be $O(3\lambda^2)$ which is
also correct).
It is clear that $dp=<\psi^*\psi>d\lambda$ cannot hold globally since
$<\psi^*\psi>$ should have poles at $D+D^*$ and the correct global 
statement follows from \cite{ka,kt}, namely $(dp/<\psi^*\psi>)=d\hat
{\Omega}$ where $d\hat{\Omega}$ is the unique meromorphic differential 
with a double pole at $\infty$ and zeros at the poles of $\psi^*\psi$ (i.e.
at $D+D^*$); this characterizes $d\hat{\Omega}$ and one can state
\\[3mm]\indent {\bf THEOREM 3.2}.$\,\,$  With 
the notations as above $dp = <\psi\psi^*>d\hat{\Omega},\,\,
dE\sim -<\psi^*L^1\psi>d\hat{\Omega}$, and
$d\Omega \sim -<\psi^*A^1\psi>d\hat{\Omega}$.
\\[3mm]\indent
{\bf REMARK 3.3.}$\,\,$ From (\ref{NS}) one obtains
now $\partial_T<\gamma>=-\partial_X<\hat{{\cal L}}>$.  Further since
$p_T=\Omega_X$ corresponds to $\partial_Tdp=\partial_Xd\Omega$ and 
$\partial_Tdp\sim\partial_T<\gamma>d\hat{\Omega}$ (note $d\hat{\Omega}$
will not depend on slow variables), one can formally write
\be
-\partial_X<\hat{{\cal L}}>\sim\partial_T<\gamma>\sim\frac{\partial_Tdp}
{d\hat{\Omega}}\sim \frac{\partial_Xd\Omega}{d\hat{\Omega}}
\label{PFF}
\ee
suggesting that
\be
<\hat{{\cal L}}>d\hat{\Omega}\sim-d\Omega\sim<\psi^*A^1\psi>d\hat{\Omega}\,
\Rightarrow
\,\,<\hat{{\cal L}}>\sim <\psi^*A^1\psi>
\label{PGG}
\ee
This is consistent with (\ref{NS}) since
\be
\partial_T<\gamma>+\partial_X<\hat{{\cal L}}>\sim
\partial_T\left(\frac{dp}{d\hat{\Omega}}\right)-\partial_X\left(
\frac{d\Omega}{d\hat{\Omega}}\right)=0
\label{PHH}
\ee
via $\partial_Tdp=\partial_Xd\Omega$.
\\[3mm]\indent
In summary we can also state
\\[3mm]\indent {\bf THEOREM 3.4}.$\,\,$ 
The quantity $\psi\psi^*$ is seen to determine the Whitham hierarchy a;nd
the differentials $dp,\,\,dE,\,\,d\Omega$, etc.
\\[3mm]\indent {\bf REMARK 3.5}.$\,\,$  Since $D+D^*-2\infty\sim K_{\Sigma},
\,\,\psi\psi^*$ is determined by a section of $K_{\Sigma}$ (global
point of view) but it is relations based on $<\psi^*L^1\psi>,\,\,
<\psi^*A^1\psi>,$ etc. (based on the Krichever averaging process)
which reveal the ``guts" of $\psi\psi^*$ needed for averaging and the
expression of differentials.

\section{DISPERSIONLESS THEORY}
\renewcommand{\theequation}{4.\arabic{equation}}\setcounter{equation}{0}

\subsection{General framework for KP}
We give next a brief sketch of some ideas regarding dispersionless KP
(dKP) following mainly \cite{ch,ci,cl,gh,kn,ta} to which we refer for 
philosophy.  We will make various notational adjustments as we go along.  One
can think of fast and slow variables with $\epsilon x=X$ and $\epsilon t_n=
T_n$ so that $\partial_n\to\epsilon\partial/\partial T_n$ and $u(x,t_n)
\to\tilde{u}(X,T_n)$ to obtain from the KP equation $(1/4)u_{xxx}+3uu_x
+(3/4)\partial^{-1}\partial^2_2u=0$ the equation $\partial_T\tilde{u}=3
\tilde{u}\partial_X\tilde{u}+(3/4)\partial^{-1}(\partial^2\tilde{u}/
\partial T_2^2)$ when $\epsilon\to 0$ ($\partial^{-1}\to(1/\epsilon)
\partial^{-1}$).  In terms of hierarchies the theory can be built around the
pair $(L,M)$ in the spirit of \cite{cd,ci,ta}.  Thus writing $(t_n)$ for
$(x,t_n)$ (i.e. $x\sim t_1$ here) consider
\be
L_{\epsilon}=\epsilon\partial+\sum_1^{\infty} u_{n+1}(\epsilon,T)
(\epsilon\partial)^{-n};\,\,M_{\epsilon}=\sum_1^{\infty}nT_nL^{n-1}_{\epsilon}
+\sum_1^{\infty}v_{n+1}(\epsilon,T)L_{\epsilon}^{-n-1}
\label{YAA}
\ee
Here $L$ is the Lax operator $L=\partial+\sum_1^{\infty}u_{n+1}\partial^{-n}$
and $M$ is the Orlov-Schulman operator defined via $\psi_{\lambda}=M\psi$.
Now one assumes $u_n(\epsilon,T)=U_n(T)+O(\epsilon)$, etc. and 
set (recall $L\psi=\lambda\psi$)
$$
\psi=\left[1+O\left(\frac{1}{\lambda}\right)\right]exp
\left(\sum_1^{\infty}\frac{T_n}
{\epsilon}\lambda^n\right)=exp\left(\frac{1}
{\epsilon}S(T,\lambda)+O(1)\right);$$
\be
\tau=exp\left(\frac{1}{\epsilon^2}F(T)+O\left(\frac{1}{\epsilon}\right)
\right)
\label{YBB}
\ee
We recall that $\partial_nL=[B_n,L],\,\,B_n=L^n_{+},\,\,\partial_nM
=[B_n,M],\,\,[L,M]=1,\,\,L\psi=\lambda\psi,\,\,\partial_{\lambda}\psi
=M\psi,$ and $\psi=\tau(T-(1/n\lambda^n))exp[\sum_1^{\infty}T_n\lambda^n]/
\tau(T)$.  Putting in the $\epsilon$ and using $\partial_n$ for
$\partial/\partial T_n$ now, with $P=S_X$, one obtains
\be
\lambda=P+\sum_1^{\infty}U_{n+1}P^{-n};\,\,
P=\lambda-\sum_1^{\infty}P_i\lambda^{-1};
\label{YCC}
\ee
$${\cal
M}=\sum_1^{\infty}nT_n\lambda^{n-1}+\sum_1^{\infty}V_{n+1}\lambda^{-n-1};
\,\,\partial_nS={\cal B}_n(P)\Rightarrow \partial_nP=\hat{\partial}
{\cal B}_n(P)$$
where $\hat{\partial}\sim \partial_X+(\partial P/\partial X)\partial_P$
and $M\to {\cal M}$.
Note that one assumes also $v_{i+1}(\epsilon,T)=V_{i+1}(T)+O(\epsilon)$; 
further
for $B_n=\sum_0^nb_{nm}\partial^m$ one has ${\cal B}_n=\sum_0^nb_{nm}
P^m$ (note also $B_n=L^n+\sum_1^{\infty}\sigma_j^nL^{-j}$).  
We list a few additional formulas which are 
easily obtained (cf. \cite{ci}); thus, writing $\{A,B\}=\partial_PA\partial A
-\partial A\partial_PB$ one has
\be
\partial_n\lambda=\{{\cal B}_n,\lambda\};\,\,\partial_n{\cal M}
=\{{\cal B}_n,{\cal M}\};\,\,\{\lambda,{\cal M}\}=1
\label{YDDD}
\ee
Now we can write $S=\sum_1^{\infty}T_n\lambda^n+\sum_1^{\infty}S_{j+1}
\lambda^{-j}$ with $\partial_mS_{j+1}=\tilde{\sigma}_j^m,\,\,V_{n+1}=-nS_{n+1}$,
and $\partial_{\lambda}S={\cal M}\,\, (
\sigma_j^m\to\tilde{\sigma}_
j^m$).  Further 
\be
{\cal B}_n=\lambda^n+\sum_1^{\infty}\partial_nS_{j+1}\lambda^{-j};\,\,
\partial S_{n+1}\sim -P_n\sim -\frac{\partial V_{n+1}}{n}\sim
-\frac{\partial\partial_n F}{n}
\label{YEEE}
\ee
\indent
We sketch next a few formulas from \cite{kn} (cf. also \cite{kq,kr}).  First
it will be important to rescale the $T_n$ variables and write 
$t'=nt_n,\,\,T_n'=nT_n,\,\,
\partial_n=n\partial'_n=n(\partial/\partial T'_n)$.  Then
\be
\partial'_nS=\frac{\lambda^n_{+}}{n};\,\,\partial'_n\lambda=\{{\cal Q}_n,
\lambda\}\,\,({\cal Q}_n=\frac{{\cal B}_n}{n});
\label{YFF}
\ee
$$\partial'_nP=\hat{\partial}{\cal Q}_n=\partial{\cal Q}_n+\partial_P
{\cal Q}_n\partial P;\,\,\partial'_n{\cal Q}_m-\partial'_m{\cal Q}_n=
\{{\cal Q}_n,{\cal Q}_m\}$$
Now think of $(P,X,T'_n),\,\,n\geq 2,$ as basic Hamiltonian variables
with $P=P(X,T'_n)$.  Then $-{\cal Q}_n(P,X,T'_n)$ will serve as a
Hamiltonian via
\be
\dot{P}'_n=\frac{dP'}{dT'_n}=\partial{\cal Q}_n;\,\,\dot{X}'_n=\frac
{dX}{dT'_n}=-\partial_P{\cal Q}_n
\label{YGG}
\ee
(recall the classical theory for variables $(q,p)$ involves $\dot{q}=
\partial H/\partial p$ and $\dot{p}=-\partial H/\partial q$).  The function
$S(\lambda,X,T_n)$ plays the role of part of a generating function 
$\tilde{S}$ for the Hamilton-Jacobi theory with action angle variables 
$(\lambda,-\xi)$ where
\be
PdX+{\cal Q}_ndT'_n=-\xi d\lambda-K_ndT'_n+d\tilde{S};\,\,K_n=-R_n=-
\frac{\lambda^n}{n};
\label{YHH}
\ee
$$\frac{d\lambda}{dT'_n}=\dot{\lambda}'_n=\partial_{\xi}R_n=0;\,\,\frac
{d\xi}{dT'_n}=\dot{\xi}'_n=-\partial_{\lambda}R_n=-\lambda^{n-1}$$
(note that $\dot{\lambda}'_n=0\sim\partial'_n\lambda=\{{\cal Q}_n,
\lambda\}$).  To see how all this fits together we write
\be
\frac{dP}{dT'_n}=\partial'_nP+\frac{\partial P}{\partial X}\frac{dX}
{dT'_n}=\hat{\partial}{\cal Q}_n+\frac{\partial P}{\partial X}\dot
{X_n}'=\partial{\cal Q}_n+\partial P\partial_P{\cal Q}_n+\partial P
\dot{X}'_n
\label{YI}
\ee
This is compatible with (\ref{YGG}) and Hamiltonians $-{\cal Q}_n$.  Furthermore
one wants
\be
\tilde{S}_{\lambda}=\xi;\,\,\tilde{S}_X=P;\,\,\partial'_n\tilde{S}=
{\cal Q}_n-R_n
\label{YJJ}
\ee
and from (\ref{YHH}) one has
\be
PdX+{\cal Q}_ndT'_n=-\xi d\lambda+R_ndT'_n+\tilde{S}_XdX+\tilde{S}_
{\lambda}d\lambda+\partial'_n\tilde{S}dT'_n
\label{YKK}
\ee
which checks.  We note that $\partial'_nS={\cal Q}_n={\cal B}_n/n$ and
$S_X=P$ by constructions and definitions.  Consider $\tilde{S}=S-\sum_2^{\infty}
\lambda^nT'_n/n$.  Then $\tilde{S}_X=S_X=P$ and $\tilde{S}_n'=S_n'-R_n=
{\cal Q}_n-R_n$ as desired with $\xi=\tilde{S}_{\lambda}=S_{\lambda}-
\sum_2^{\infty}T'_n\lambda^{n-1}$.  It follows that
$\xi\sim{\cal M}-\sum_2^{\infty}T'_n\lambda^{n-1}=X+\sum_1^{\infty}V_{i+1}
\lambda^{-i-1}$.  If $W$ is the gauge operator such that $L=W\partial W^{-1}$
one sees easily that
\be
M\psi
=W\left(\sum_1^{\infty}kx_k\partial^{k-1}\right)W^{-1}\psi=\left(G+
\sum_2^{\infty}kx_k\lambda^{k-1}\right)\psi
\label{YLL}
\ee
from which follows that $G=WxW^{-1}\to\xi$.  This shows that $G$ is
a very fundamental object and this is encountered in various places
in the general theory (cf. \cite{cd,ci,se,ye}).
\\[3mm]\indent {\bf REMARK 4.1.}$\,\,$ We refer here also to \cite{ch,cl}
for a complete characterization of dKP and the solution of the dispersionless
Hirota equations.

\subsection{Dispersonless theory for KdV}
We give a special treatment here for KdV since it is needed in
\cite{cp,cs}.  Thus
following \cite{ca,ce,ci,da} we write
\be
L^2=L^2_{+}=\partial^2+q=\partial^2-u\,\,(q=-u=2u_2);\,\,q_t-6qq_x-q_{xxx}=0;
\label{YMM}
\ee
$$B=4\partial^3+6q\partial+3q_x;\,\,L^2_t=[B,L^2];\,\,q=-v^2-v_x\sim
v_t+6v^2v_x+v_{xxx}=0$$
($v$ satisfies the mKdV equation).  KdkV is Galilean invariant ($x'=x-
6\lambda t,\,\,t'=t,\,\,u'=u+\lambda$) and consequently one can consider
$L+\partial^2+q-\lambda=(\partial+v)(\partial-v,\,\,q-\lambda=-v_x-v^2,\,\,
v=\psi_x/\psi,$ and $-\psi_{xx}/\psi=q-\lambda$ or $\psi_{xx}+q\psi=\lambda\psi$
(with $u'=u+\lambda\sim q'=q-\lambda$).  The $v$ equation in (\ref{YMM}) becomes
then $v_t=\partial(-6\lambda v+2v^3-v_{xx})$ and for $\lambda=-k^2$ one
expands for $\Im k>0,\,\,|k|\to\infty$ to get
$(\bullet\bullet)\,\,
v\sim ik+\sum_1^{\infty}(v_n/(ik)^n)$.  The $v_n$ are conserved
densities and with $2-\lambda=-v_x-v^2$ one obtains
\be
p=-2v_1;\,\,2v_{n+1}=-\sum_1^{n-1}v_{n-m}v_m-v'_n;\,\,2v_2=-v'_1
\label{YNN}
\ee
Next for $\psi''-u\psi=-k^2\psi$
write $\psi_{\pm}\sim exp(\pm ikx)$ as $x\to\pm\infty$.  Recall also the 
transmission and reflection coefficient formulas (cf. \cite{ca})
$T(k)\psi_{-}=R(k)\psi_{+}+\psi_{+}(-k,x)$ and $T\psi_{+}=R_L\psi_{-}+
\psi_{-}(-k,x)$.  Writing e.g. $\psi_{+}=exp(ikx+\phi(k,x))$ with
$\phi(k,\infty)=0$ one has $\phi''+2ik\phi'+(\phi')^2=u$.  Then
$\psi'_{+}/\psi_{+}=ik+\phi'=v$ with $q-\lambda=-v_x-v^2$.  Take then
\be
\phi'=\sum_1^{\infty}\frac{\phi_n}{(2ik)^n};\,\,v\sim ik+\phi'=ik+
\sum\frac{v_n}{(ik)^n}\Rightarrow\phi_n=2^nv_n
\label{YOO}
\ee
Furthermore one knows (cf. \cite{ca})
\be
log T=-\sum_0^{\infty}\frac{c_{2n+1}}{k^{2n+1}};\,\,c_{2n+1}=\frac{1}{2\pi i}
\int_{-\infty}^{\infty}k^{2n}log(1-|R|^2)dk
\label{YPP}
\ee
(assuming for convenience that there are no bound states).  Now for
$c_{22}=R_L/T$ and $c_{21}=1/T$ one has as $k\to -\infty\,\,(\Im k>0)$ the
behavior $\psi_{+}exp(-ikx)\to c_{22}exp(-2ikx)+c_{21}\to c_{21}$.  Hence
$exp(\phi)\to c_{21}$ as $x\to -\infty$ or $\phi(k,-\infty)=-log T$ which
implies
\be
\int_{-\infty}^{\infty}\phi'dx=log T=\sum_1^{\infty}\int_{-\infty}^{\infty}
\frac{\phi_ndx}{(2ik)^n}
\label{YQQ}
\ee
Hence $\int\phi_{2m}dx=0$ and $c_{2m+1}=-\int\phi_{2m+1}dx/(2i)^{2m+1}$.
The $c_{2n+1}$ are related to Hamiltonians $H_{2n+1}=\alpha_nc_{2n+1}$
as in \cite{cd,ce} and thus the conserved densities $v_n\sim \phi_n$ give
rise to Hamiltonians $H_n$ (n odd).  There are action angle variables
$P=klog|T|$ and $Q=\gamma arg(R_L/T)$ with Poisson structure $\{F,G\}\sim
\int(\delta F/\delta u)\partial (\delta G/\delta u)dx$ (we omit the
second Poisson structure here).  
\\[3mm]\indent
Now look at the dispersionless theory based on $k$ where $\lambda
\sim(ik)^2=-k^2$.  One obtains for $P=S_X,\,\,P^2+q=-k^2$, and we write
${\cal P}=(1/2)P^2+p=(1/2)(ik)^2$ with $q\sim 2p\sim 2u_2$.  One has
$\partial k/\partial T_{2n}=\{(ik)^{2n},k\}=0$ and from $ik=P(1+qP^{-2})^
{1/2}$ we obtain
\be
ik=P\left(1+\sum_1^{\infty}{\frac{1}{2}\choose m}q^mP^{-2m}\right)
\label{YRR}
\ee
(cf. (\ref{YCC}) with $u_2=q/2$).  The flow equations become then
\be
\partial'_{2n+1}P=\hat{\partial}{\cal Q}_{2n+1};\,\,\partial'_{2n+1}(ik)
=\{{\cal Q}_{2n+1},ik\}
\label{YSS}
\ee
Note here some rescaling is needed since we want $(\partial^2+q)^{3/2}_{+}
\partial^3+(3/2)q\partial+(3/4)q_x=B_3$ instead of our previous $B_3\sim
4\partial^3 +6q\partial+3q_x$.  Thus we want ${\cal Q}_3=(1/3)P^3+(1/2)qP$
to fit the earlier notation.  The Gelfand-Dickey resolvant 
coefficients are defined via $R_s(u)=(1/2)Res(\partial^2-u)^{s-(1/2)}$
and in the dispersionless picture 
$R_s(u)\to (1/2)r_{s-1}(-u/2)$ (cf. \cite{ci}) where
$$
r_n=Res(-k^2)^{n+(1/2)}=\left(
\begin{array}{c}
n+(1/2)\\
n+1
\end{array}\right)q^{n+1}=
\frac{(n+1/2)\cdots(1/2}{(n+1)!}q^{n+1};$$
\be
\,\,2\partial_qr_n=(2n+1)r_{n-1}
\label{YTT}
\ee
The inversion formula corresponding to (\ref{YCC}) is $P=ik-\sum_1^{\infty}
P_j(ik)^{-j}$ and one can write
\be
\partial'_{2n+1}(P^2+q)=\partial'_{2n+1}(-k^2);\,\,\partial'_{2n+1}q
=\frac{2}{2n+1}\partial r_n=\frac{2}{2n+1}\partial_qr_nq_X=
q_Xr_{n-1}
\label{YUU}
\ee
Note for example $r_0=q/2,\,\,r_1=3q^2/8,\,\,r_2=5q^3/16,\cdots$ and
$\partial'_Tq=q_Xr_0=(1/2)qq_X$ (scaling is needed in (\ref{YMM}) here for
comparison).  Some further calculation gives for $P=ik-\sum_1^{\infty}
P_n(ik)^{-n}$ 
\be
P_n\sim-v_n\sim-\frac{\phi_n}{2^n};\,\,c_{2n+1}=(-1)^{n+1}\int_
{-\infty}^{\infty}P_{2n+1}(X)dX
\label{YVV}
\ee
The development above actually gives a connection between inverse
scattering and the dKdV theory (cf. \cite{ch,ci,cl} for more on this).

\section{WHITHAM EQUATIONS AND SYMPLECTIC FORMS}
\renewcommand{\theequation}{5.\arabic{equation}}\setcounter{equation}{0}

\subsection{General relations}

In \cite{kc,kj,kl,kt} there is a development of Whitham theory and
symplectic forms in a general context (cf. also \cite{cn,co,cp,cs,
dq,di,dj,dk}).
The idea is to have a universal Whitham hierarchy based on a space
$\hat{M}_{gN}$ of smooth algebraic curves $\Sigma_g$ with $N$ punctures
$P_{\alpha},\,\,\alpha=1,\cdots,N$, and local coordinates $k_{\alpha}^{-1}(P)$
(here $P$ is a point on $\Sigma_g$).  For each such datum and a set of 
$g$ points $\gamma_1,\cdots,\gamma_g\in\Sigma_g$ in general position (or
equivalently a point in the Jacobian $J(\Sigma_g)$) standard 
algebro-geometric constructions give a quasiperiodic solution of some 
integrable nonlinear PDE.  For KP one has $N=1$ and for Toda $N=2$; we
concentrate here on $N=1$. Note that $P$ is used to represent
a point on $\Sigma_g$ 
(corresponding to $P=S_X$ in dispersionless theory), with $dp
=d\Omega_1$ and $p(P)=\int^Pdp$.  From now
on we will be scrupulous in distinguishing $d\Omega_i$ and $\Omega_i
=\int^Pd\Omega_i$.
In \cite{kc,kj} one uses $k_1^{-1}$ as a local coordinate at $\infty$
and we will think here of $k_1\sim k\sim\lambda$ where $\lambda$
corresponds to the ``spectral" variable for KP (which in dKP becomes an
action variable).  Further $p(k,T)$ is specified as $\Omega_1(k,T)$ 
so $k$ and $P$ both represent points on $\Sigma_g$.
Thus one must be careful in interperting $P=S_X$ from dKP when working
on $\Sigma_g$.  In fact the $\hat{S}$ of (\ref{AE}) satisfies $\hat{S}_X=U$
as indicated before (\ref{AT})
and we will see below that
there will be an action term ${\cal S}(p)=\sum T_i\Omega_i(p)$ with 
$\partial_i{\cal S}=\Omega_i$.  Hence for $\Omega_1=p$ one has $\partial_X
{\cal S}=p$ and this plays the role of $P=S_X$ 
(from dispersionless theory) on $\Sigma_g$.
Thus, for simplicity, in what follows $P\in\Sigma_g,\,\,k\sim P_1
\sim\infty$ with no other punctures, and $p\sim\Omega_1$.
\\[3mm]\indent
Now in \cite{kc,kj} one relates the Whitham equations 
$(\spadesuit\spadesuit)\,\,\partial_i\Omega_j=
\partial_j\Omega_i\,\,(\partial_i\sim\partial/\partial T_i)$,
where the $\Omega_i$ are expressed in suitable variables
as indicated below, to zero
curvature equations
\be
\partial_A\Omega_B-\partial_B\Omega_A+\{\Omega_A,\Omega_B\}=0;\,\,
\{f,g\}=f_Xg_p-f_pg_X
\label{O}
\ee
(note the order of terms in $\{f,g\}$ is different from (\ref{YDD})).  Here
the $A,\,B$ correspond to $(1,i),\,\,1\sim P_1,$ and $i\sim\Omega_i$ where
one thinks of $k=k(p,T)$ or $p=p(k,T)$ with $T_1=X,\,\,\Omega_A=\Omega_A
(p,T),$ and
\be
\Omega_i=\int^{P\sim k}d\Omega_i=k^i+O(k^{-1});\,\,k^i(p)=\sum_1^iw_{is}
p^s
\label{N}
\ee
The analogy here is to ${\cal B}_i$ of (\ref{YEEE}) for example
where
\be
{\cal B}_n=\lambda^n_{+}=\lambda^n-\sum_1^{\infty}\frac{F_{nm}}{m}\lambda^
{-m}=\sum_0^nb_{nm}P^m
\label{M}
\ee
Note that $b_{n0}\not= 0$ is possible however and recall that some 
normalization ($\oint_{A_j}d\Omega_i=0$ for example) 
will hold; in any case the 
development is parallel.  Now if $E$ is an arbitrary function of $(p,T)$ one
can regard (\ref{O}) as compatibility conditions for
\be
\partial_AE = \{E,\Omega_A\}
\label{CH}
\ee
(easy exercise, using the
Jacobi identity) and this places one in the context of what are called
algebraic solutions (when there is a global solution of (\ref{CH}).  Here
global means that $E$ is a meromorphic solution of (\ref{CH}) such that
$\{E(p,T),k(p,T)\}=0$ so there exists $f(E)$ with $k(p,T)=f(E(p,T))$
(note then $\{E,k\}$ becomes $E_Xf'E_p-E_pf'E_X=0$).  Generally one will
stipulate that $dE$ be a normalized meromorphic differential of the second
kind. 
When $\partial_pE\not= 0$ one can write $p=p(E,T)$
and $\partial_Af(p,T) = \partial_AF(E,T) + (\partial F/\partial E)
\partial_AE\,\,(F(E,T) = f(p,T)$).  Then for such an $E$ (\ref{O})
becomes
\be
\partial_A\Omega_B(E,T) = \partial_B\Omega_A(E,T)
\label{CI}
\ee
corresponding to $(\spadesuit\spadesuit)$.
Indeed we note that ($\Omega'_A\sim (\partial\Omega_A/\partial E),\,\,
E'\sim(\partial E/\partial p)$)
\be
(\ref{O}) = \partial_A\Omega_B + \Omega'_BE_A -\partial_B\Omega_A -
\Omega'_AE_B + (\partial_X\Omega_A +\Omega'_AE_X)\Omega'_BE' -
\label{CJ}
\ee
$$- \Omega'_AE'(\partial_X\Omega_B + \Omega'_BE_X) = \partial_A\Omega_B
-\partial_B\Omega_A + $$
$$ +\Omega'_B(E_A+E'\partial_X\Omega_A) -
\Omega'_A(E_B + E'\partial_X\Omega_B) = 0$$
But (\ref{CH}) implies for example that $E_A = E_X\partial_p\Omega_A(p,T) -
E'\partial_X\Omega_A(p,T) = E_X\Omega'_AE' - E'(\partial_X\Omega_A(E,T)
+ \Omega'_AE_X) = -E'\partial_X\Omega_A\Rightarrow E_A + E'\partial_X
\Omega_A(E,T) = 0$.  Similarly $E_B + E'\partial_X\Omega_B(E,T) = 0$ and
(\ref{CJ}) implies (\ref{CI}).  
Thence, via (\ref{CI}), in coordinates $E,T$
we can introduce a ``potential" ${\cal S}(E,T)$ such that
\be
\Omega_A(E,T) =
\partial_A{\cal S}(E,T)
\label{L}
\ee
and define
\be
\omega=\sum \Omega_AdT_A
\label{K}
\ee
(note this ${\cal S}$ is not a priori related to the $\hat{S}$ of (\ref{AE})).
Then
for $Q = \partial {\cal S}/\partial E$ and $\delta\sim$ full exterior derivative
\be
\delta {\cal S}(E,T) - QdE = \partial_E{\cal S}dE + \sum\partial_A
{\cal S}dT_A -QdE
= \sum \Omega_AdT_A=\omega
\label{CK}
\ee
(note also that $p = \partial {\cal S}/\partial X = \Omega_1$).  It follows
that $\delta\omega = \sum\delta\Omega_A\wedge dT_A$ formally. 
Note that we are treating here $dE$ and $dT_A$ as independent objects
and in this spirit from $\partial_A{\cal S}=\Omega_A$ and 
$d{\cal S}=QdE$ one obtains
$d\Omega_A=d(\partial_A{\cal S})=\partial_A(QdE)=(\partial_AQ)dE$ so that
\be
\partial_AQ=\frac{d\Omega_A}{dE};\,\,\partial_XQ=\frac{dp}{dE}
\label{J}
\ee
(thus $\partial_AdE=0$ in this context - i.e. $\partial_AdE\not= d(\partial_A
E)$).  In the same spirit from $\omega=\sum\partial_A{\cal S}dT_A=\delta 
{\cal S}-
{\cal S}_EdE=\delta {\cal S}-QdE$ we get
\be
\delta\omega=\delta^2{\cal S}-\delta Q\wedge dE=dE\wedge\delta Q
\label{I}
\ee
This is set equal to $\delta E\wedge\delta Q$ in \cite{kc,kj} but this seems
incorrect unless other assumptions are made, so a justification
follows.  Thus first
for a more complete discussion of related manipulation we refer to \cite{ta}
where ($S\sim{\cal S}$ generically now)
\be
{\cal M}=\partial_{\lambda}{\cal S}\sim Q=\partial_E{\cal S};
\,\,{\cal B}_n=\partial_n{\cal S}
\sim\Omega_A=\partial_A{\cal S};
\label{H}
\ee
$$d{\cal S}={\cal M}d\lambda+\sum 
{\cal B}_ndT_n\sim \delta {\cal S}=QdE+\sum\Omega_AdT_A$$
We note next that, using (\ref{J}) and $\partial_Af=\partial_AF+
F_E\partial_AE$ before (\ref{CI}),
\be
\{Q(p,T),E(p,T)\}=\partial_XQ\partial_pE-\partial_pQ\partial_XE=
\label{G}
\ee
$$=(\partial_XQ(E,T)+Q_EE_X)E_p-Q_EE_pE_X=Q_XE_p=\frac{dp}{dE}\frac{dE}
{dp}=1$$
(analogous to $\{\lambda,{\cal M}\}=1$ in \cite{ta}, where $\{\lambda,
{\cal M}\}=\lambda_P{\cal M}_X-\lambda_X{\cal M}_P$).  Further note
from $\{Q,E\}=1,\,\,\partial_AE=\{E,\Omega_A\}$, and the Jacobi identity
one has
\be
\{\partial_AQ,E\}+\{Q,\partial_AE\}=0\Rightarrow \{\partial_AQ,E\}+\{\{
E,\Omega_A\},Q\}=
\label{E}
\ee
$$=\{\partial_AQ,E\}-\{\{Q,\Omega_A\},E\}$$
since $\{\{Q,E\},\Omega_A\}=0$.  Then from $\{f,E\}=0$ one gets
$f_XE_p-f_pE_X=0$ so for $f(p,T)=F(E,T)$ one has 
$(F_X+F_EE_X)E_p-F_EE_pE_X=F_XE_p=0$
($F\sim\partial_AQ-\{Q,\Omega_A\}$ here).
Thus up to functions of $T_n,\,\,n>1$ 
there results 
\be
\partial_AQ=\{Q,\Omega_A\}
\label{D}
\ee
which is not quite satisfactory.  A better approach follows from \cite{ta}.
Thus if we assume $\delta\omega=\delta E\wedge\delta Q$ one can write out
$\delta\omega=\delta E\wedge\delta Q=\sum\delta\Omega_A\wedge dT_A$ to get
\be
\sum_A(\partial_p\Omega_Adp+\sum\partial_B\Omega_AdT_B)\wedge
dT_A=
\label{C}
\ee
$$=(E_pdp+\sum\partial_CEdT_C)\wedge(Q_pdp+\sum_D\partial_DQdT_D)
$$
The coefficients of $dp\wedge dT_A$ and $dX\wedge dT_A$ are then
\be
\begin{array}{c}
\partial_p\Omega_A=E_p\partial_AQ-Q_p\partial_AE;\\
\partial_X\Omega_A=\partial_XE\partial_AQ-\partial_AE\partial_XQ
\end{array}
\label{A}
\ee
Solving for $\partial_AQ$ gives (\ref{D}).  Hence if we stipulate that
(\ref{D}) holds then $\delta\omega=\delta E\wedge\delta Q$ is permitted
in (\ref{I}).

\subsection{Times, coordinates, and differentials}

Another step needing further consideration is the construction 
${\cal S}(p,T)=\int^pQ(p,T)dE(p,T)$ or 
${\cal S}(p,T)=\sum(T_A-T_A^0)\Omega_A
(p,T)$.  This works for the genus zero situation with 
$\partial_A{\cal S}=\Omega_A$ (but this last 
formula needs proving - cf. \cite{kj}
for a ``proof" when $E\sim\lambda^n=\lambda^n_{+},\,\,\lambda\sim k$).
We remark also that for any formal
power series $Q(p)=\sum_1^{\infty}b_jp^j$ one can define times $\hat{T}_i$ via
\be
\hat{T}_i = \frac{1}{i}Res_{p=\infty}k^{-i}(p)Q(p)dE(p)
\label{CS}
\ee
Thus for $Q\sim (d{\cal S}/dE)$ with 
${\cal S}(p)=\sum_1^{\infty}T_i\Omega_i\sim\sum_1^
{\infty}T_ik^i+O(k^{-1})$
one has 
\be
\hat{T}_i = \frac{1}{i}Res_{k=\infty}k^{-i}d{\cal S} = \frac{1}{i}Res_{k=\infty}
\sum_1^{\infty}jT_jk^{j-i-1} = T_i
\label{CT}
\ee
\indent
For the general situation without additional punctures we will describe
matters as in \cite{co} following \cite{kt} (cf. also \cite{di,dj,kc,kj}).
Thus take e.g.
\be
E=p^n+u_{n-2}p^{n-2}+\cdots+u_0+O(p^{-1})\,\,or\,\,E\sim k^n+O(k^{-1})
\label{CU}
\ee
This includes $E\sim\lambda^n_{+}$ or $E\sim\lambda^n$ or $dE\sim
d\Omega_n$ (when normalized).  For Whitham equations we will want
normalization such as $\oint_{A_i}dE=0$ but a larger theory can also
be envisioned as in \cite{kt}.  Let us think of $\Re\oint_{A_i}d\Omega_n
=0=\Re\oint_{B_i}d\Omega_n$ here.
Similarly take
\be
dQ=d(k^m+O(k^{-1}))
\label{CUU}
\ee
so $dQ\sim d\Omega_m$ is appropriate when normalized.  If the $dE$ and $dQ$
are not normalized there are $g$ free parameters for each, corresponding
to the holomorphic differentials which can be added without affecting 
asymptotic behavior.  Look at the configuration space ${\cal M}_g(n,m)=
\{\Sigma_g,P_1\sim\infty,[k]_n,[k]_m,E,Q\}$.  Note that an n-jet of
coordinates $\sum_0^nk^ja_j$ is an equivalence class via $k'=k+O(k^{-n-1})$
and we will specify e.g. $Q=\sum_{-\infty}^mc_jk^j$ with no logarithmic 
term; the corresponding jet $[k]_m$ is then determined by $m+1$ coefficients
$c_0,\cdots,c_m$ plus $g$ parameters for the holomorphic differentials which
could be added without modifying the asymptotics.  Hence in ${\cal M}_g
(n,m)$ we will have $3g-3+1$ parameters (for the moduli space of Riemann
surfaces with genus $g$ with one puncture), plus $n+m+2+2g$ parameters
for $E,\,Q$; hence there are $5g+n+m$ parameters involved in specifying
${\cal M}_g(n,m)$.  This number will be matched by $5g+n+m$ ``times"
\be
T_j=\frac{1}{j}Res_{\infty}(k^{-j}QdE),\,\,(j=1,\cdots,n+m);\,\,
\tau_{A_i,E}=\oint_{A_i}dE;
\label{GC}
\ee
$$\tau_{B_i,E}=\oint_{B_i}dE;\,\,
\tau_{A_i,Q}=\oint_{A_i}dQ;\,\,
\tau_{B_i,Q}=\oint_{B_i}dQ;\,\,a_i=\oint_{A_i}QdE$$
where $i=1,\cdots,g$.
\\[3mm]\indent
Now let ${\cal D}$ be the open set in ${\cal M}_g(n,m)$ where the zero
divisors of $dE$ and $dQ$ do not intersect (i.e. the sets $\{\gamma;\,\,
dE(\gamma)=0\}$ and $\{\gamma;\,\,dQ(\gamma)=0\}$ do not intersect).
It is proved in \cite{kt} that near each point in 
${\cal D}$ the $5g+n+m$ functions
$T_k,\,\,\tau_{A_i,E},\,\,\tau_{B_i,E},\,\,
\tau_{A_i,Q},\,\,\tau_{B_i,Q},\,\,a_i$ have linearly independent differentials,
and thus define a local holomorphic coordinate system.  Further the joint
level sets of these functions (omitting the $a_i$) define a smooth
g-dimensional foliation of ${\cal D}$, independent of the choices made in
defining the coordinates themselves.
Then ${\cal M}_g={\cal M}_g(n,m)$ can be taken as a base space for two
fibrations ${\cal N}^g$ and ${\cal N};\,\,{\cal N}^g$ has fiber
$S^g(\Sigma)\simeq J(\Sigma)=$ Jacobian variety (via the Abel map
$(\gamma_1,\cdots,\gamma_g)\to \sum_{i=1}^g\int_P^{\gamma_i}d\omega_j)$
and ${\cal N}$ has fiber $\Sigma$ (all over a point $(\Sigma,P,[k]_n,[k]_m
E,Q)\in{\cal M}_g$).  We consider only leaves ${\cal M}$ of the foliation
of ${\cal D}$ indicated above and look at the fibration ${\cal N}$ or
${\cal N}^g$ over the base ${\cal M}$.
First, although $E$ and $Q$ are multivalued on
the universal fibration their differentials are well defined on ${\cal N}$.
The idea here is that $E$ and $Q$ are well defined near $P_1$ and their
analytic continuations by different paths can only change by multiples
of their residues or periods along closed cycles.  But on a leaf of the
foliation the ambiguities remain constant and disappear upon differentiation.
Hence one has differentials $\delta E$ and $\delta Q$ on the fibrations
which reduce to $dE$ and $dQ$ acting on vectors tangent to the fiber. 
One can trivialize the fibration ${\cal N}$ with the variables
$a_1,\cdots,a_g$ along the leaf ${\cal M}$ and e.g. $E$ along the
fiber.  Then $dQ$ coincides with $(dQ/dE)dE$ where $dE$ is viewed as one
of the elements of the basis of one forms for ${\cal N}$ 
and the full differential is $\delta Q=dQ+\sum_1^g(\partial Q/
\partial a_i)da_i\equiv dQ+\delta^EQ$ (also one can take here the
constant terms in $E,\,Q$ to be zero).
This framework, with $QdE=pdE\sim\lambda_{SW}\sim$ Seiberg-Witten (SW)
differential for example leads to a symplectic form $\omega_{{\cal M}}=
\sum da_i\wedge d\omega_i$ for ${\cal N}^g$ which can be written in
various interesting ways (cf. \cite{co,cp,kt} and see also
\cite{bd,bz,dz,ea,ib,kz,kt,mf,mz,ne,sf,sc,tc});
the constructions appear in Section 5.2.
\\[3mm]\indent
In terms of Whitham theory we have differentials $d\Omega_j=d(k^j+O(k^{-1}))$
normalized as before which are coupled to the $T_j$ plus holomorphic
differentials $d\omega_j\,\,(j=1,\cdots,g$) with $\oint_{A_i}d\omega_j
=\delta_{ij}$ coupled to the $a_j$.  In addition there are differentials
$d\Omega_{E,i}$ and $d\Omega_{Q,i}$, holomorphic
on $\Sigma_g$ except for the $A_j$ cycles, where they have jumps
\be
d\Omega^{+}_{E,i} - d\Omega^{-}_{E,i} = \delta_{ij}dE;\,\,
d\Omega^{+}_{Q,i} - d\Omega^{-}_{Q,i} = \delta_{ij}dQ
\label{CY}
\ee
with normalization conditions 
\be
\oint_{A_i}d\Omega_{E,j} = \oint_{A_i}d\Omega_{Q,j} = 0
\label{CZ}
\ee
These are associated to $\tau_{B_i,E}$ and $\tau_{B_i,Q}$ and we note
that for the Whitham theory the times $\tau_{A_i,E}$ and $\tau_{A_i,Q}$
vanish by normalization of $dE$ and $dQ$.
At this point the most elegant construction of an action differential
is in \cite{ib} for Seiberg-Witten (SW) theory.  The construction there
is for a Toda theory on a hyperelliptic curve with differentials
$d\Omega_{\pm m}$ but the arguments should be adaptable to the finite zone
KP framework (see \cite{co} for a condensed version of this argument in 
\cite{ib}).  
Indeed the moduli in \cite{ib} are treated abstractly 
(they could be Casimirs for example) and no branch points are involved.
Then for the complete theory to apply it is only required that the number
$K$ of moduli $h_k$ correspond to the genus $g$.  Evidently this does not
happen generally and in the context above one would look at level sets
of the times $T_j,\,\,\tau_{B_i,E},\,\,\tau_{B_i,Q},\,\,\tau_{A_i,E},$ and
$\tau_{A_i,Q}$, leaving only $g$ variable ``times" $a_i$.  In the SW theory
many situations (e.g. $SU(n)$ susy Yang-Mills) involve constructions of
curves via flat moduli $h_k$ with $K=g$.  In such situations in \cite{ib}
one constructs by elegant and subtle arguments
\be
d{\cal S}(h_k,T)=\sum_1^ga_id\omega_i+\sum T_nd\Omega_n
\label{JB}
\ee
such that $a_i=\oint_{A_i}d{\cal S}$ with
\be
\frac{\partial d{\cal S}}{\partial T_n}=d\Omega_n;\,\,\frac{\partial d{\cal S}}
{\partial a_i}=d\omega_i
\label{JC}
\ee
There are then dual variables
$a_i^D=\oint_{B_i}d{\cal S}$ and a prepotential ${\cal F}$ such that
\be
a_i^D=\frac{\partial{\cal F}}{\partial a_i}
\label{JD}
\ee
For the introduction of ``times" $a_i$ in BA functions etc. we refer to
\cite{ne}; this will be spelled out later.
\\[3mm]\indent
In the Whitham theory sketched above we have indicated the nice features
of an ${\cal S}$ such that $Q=\partial_ES$ but the 
construction of ${\cal S}$ with 
$\partial_n{\cal S}=\Omega_n$ was left unfinished.
We refer to \cite{ib,ne} and the discussion of this in Section 6 for a more
rigorous picture but will sketch here a few useful constructions from
\cite{kc,kj,kt} for some perspective.
For the dispersionless theory there is of course no
problem (cf. Section 4 and \cite{ae,ch,ci,cl,ta}).  In \cite{kc,kj}
a number of theorems are stated (mostly without 
proof) and we will cite a few of these. 
Thus working on the subspace $D$ of $N=\{\Sigma_g,E,[k]_n,Q,[k]_m\}$
where $dQ$ and $dE$ have no common zeros one states in \cite{kc} that the 
map $D\to T=\{T_A\}$, with $T_A$ defined by (\ref{GC}) (without $\tau_{A_i,Q}$ 
or $\tau_{A_i,E}$), is nondegenerate, so that the $T_A$ determine a 
coordinate system on $D$.  The corresponding dependence of the curve
$\Sigma_g(T)$ and $dE(T)$ defines a solution of the universal Whitham
equations.  This interesting point of view considers the Whitham hierarchy
as a way to define a special system of coordinates on the moduli space
of curves with jets of local coordinates at one puncture (or more).  Next, from
\cite{kj} we have the following statement.  First define $\Pi_s$ where
$dp(\Pi_s)=0$ and set $\pi_s=p(\Pi_s)$ (note $dp$ must vanish at $2g$ points
since $deg(dp)=2g-2$ - cf. \cite{sb}).  Thus $p$ can be used as a 
coordinate except at the $2g$ points $\Pi_s$ and the parameters $\pi_s,\,\,
U_i^p=\oint_{B_i}dp$ then form a full system of local coordinates on 
$M^*=\{\Sigma_g,p,k^{-1}\}$.  The zero curvature form (\ref{O}) is then
equivalent to the compatible system of evolution equations
\be
\partial_Ak(p,T)=\{k(p,T),\Omega_A(p,T)\};\,\,\partial_AU^p_i=\partial_X
U^A_i;
\label{JE}
\ee
$$U_i^A=\oint_{B_i}d\Omega_A;\,\,\partial_A\pi_s=\partial_Ap(\Pi_s)=\partial_X
\Omega_A(\Pi_s)$$
One can add here also holomorphic differentials $d\omega_i$ as Hamiltonians
with corresponding times $a_i$ (where $dQ\sim dp$ here).
\\[3mm]\indent
If one works with the subspace $N$ of $M^*$ with fixed normalized meromorphic
differential $dE\,\,(E$ as in (\ref{CU})) there are two systems of local
coordinates.  One is given by $u_i\,\,(i=0,\cdots,n-2),\,\,\pi_s,$ and
$U^p_i$ (so $3g+n-1$ parameters whereas one expects $3g-3+1$ for Riemann
surfaces with one puncture and $n+1$ for a normalized n-jet, giving $3g+n-1$
as desired).  A second set of parameters is given by
\be
U_i^E=\oint_{B_i}dE;\,\,E_s=E(q_s)
\label{JF}
\ee
where $dE(q_s)=0\,\,(s=1,\cdots,D-g)$.  Here $dE\sim d(k^n+O(k^{-1}))$ which 
indicates a pole of order $n+1\,\,(k^{n-1}dk\sim -z^{-n+1}(dz/z^2)=
-dz/z^{n+1}$ for $z=1/k$) so $deg(dE)=2g-2\Rightarrow$ the number of zeros
is $2g-2+n+1=2g+n-1$.  For $D=3g+n-1$ we get $D-g=2g+n-1$ so $1\leq s\leq
D-g$ as in \cite{kj}.  Then the restriction of the Whitham hierarchy to $N$
is given by the compatible equations (\ref{CH}).
\\[3mm]\indent
Let now $dH_i$ be a normalized differential defined on $A_i$ such that
$\oint_{A_i}dH_i=0$.  Then there exists a unique differential $dS_H$ such
that $dS_H$ is holomorphic on $\Sigma_g$ except for the $A_i$ where it has
jumps 
\be
dS_H^{+}(P)-dS_H^{-}(P)=dH_i(P)\,\,(P\in A_i);\,\,\oint_{A_i}dS_H=0
\label{JG}
\ee
It is then stated in \cite{kj} that 
for any solution of the Whitham equations on $N$ there exists constants
$T_A^0$ and differentials
$dH_i$ independent of $T$, such that this solution is given
implicitly via
\be
\frac{d{\cal S}}{dp}(q_s,T)=0;\,\,d{\cal S}(p,T)=
\sum_A(T_A-T_A^0)d\Omega_A(p,T)+dS_H
\label{JH}
\ee
(we have added the $d$ in front of ${\cal S}$ and $\Omega_A$).
This is not entirely clear.  Then it is stated that these relations imply
$d{\cal S}=QdE$ where $Q(p)$ is holomorphic on 
$\Sigma_g$ away from the $A_i$ and
$\infty$ with jumps
\be
Q^{+}(E)-Q^{-}(E)=\frac{dH_i(E)}{dE}\,\,(E\in A_i)
\label{JI}
\ee
on the $A_i$.  This seems formally OK
and, in a situation with constant jumps $dH_i(P)=T_{Q,i}dE(P)$, 
$dQ$ is said to be single valued on $\Sigma_g$.
This seems to say $Q^{+}(E)-Q^{-}(E)=T_{Q,i}$ which corresponds
to $\oint_{B_i}dQ=-T_{Q,i}$ by (\ref{CA}) below (note one defines
the $T_{Q,i}$ differently in the following and we refer to Remark 5.1
for some clarification).
\\[3mm]\indent
Thus consider now $\tilde{N}=
\{\Sigma_g,dQ,dE\}$ with $dQ$ holomorphic away from $P_1\sim\infty$.  Take
coordinates $\pi_s,\,U_i^p,u_i,$ and the coefficients $b_{1j}$ in 
$Q=\sum_1^{\infty}b_{1j}p^j+O(p^{-1})$ and use times $T_j,\,a_j,\,T_{Q,i}=
-\oint_{B_i}dE$ and $T_{E,i}=\oint_{B_i}dQ$.  The differentials $d\Omega_
{E,i}$ and $d\Omega_{Q,i}$ coupled to $T_{E,i}$ and $T_{Q,i}$ are uniquely
defined via analyticity plus jumps on $A_i$ of the form
\be
d\Omega^{+}_{E,i}-d\Omega^{-}_{E,i}=\delta_{ij}dE;\,\,
d\Omega^{+}_{Q,i}-d\Omega^{-}_{Q,i}=\delta_{ij}dQ;
\label{JJ}
\ee
$$\oint_{A_j}d\Omega_{E,i}=\oint_{A_j}d\Omega_{Q,i}=0\,\,\,(j=1,\cdots,g)$$
As indicated earlier one can consider these times as coordinates on $\tilde
{N}$ so that $\Sigma_g=\Sigma_g(T),\,\,dQ=dQ(T),$ and $dE=dE(T)$.  Then
one asserts in \cite{kj} that for $(\clubsuit\bullet)\,\,d{\cal S}(E,T)=
Q(E,T)dE$
one has $\partial_A{\cal S}=\Omega_A$.  In fact one writes
\be
d{\cal S}=\sum_0^{\infty}T_{i}d\Omega_{i}+\sum_1^g\left(a_id\omega_i+
T_{E,j}d\Omega_{E,j}+T_{Q,j}d\Omega_{Q,j}\right)
\label{JK}
\ee
(we have added some d's and the $T_{Q,j}$ terms).  The ``proof" is marginal
at best and looks at $\partial_{Q,j}$ only.  Consider $\partial_{Q,j}
d{\cal S}(E,T)$.
From $(\clubsuit\bullet)$ it follows that $\partial_{Q,j}d{\cal S}$ is 
holomorphic
except on $A_j$.  On different sides of $A_j$ the coordinates are $E^{-}$
and $E^{+}=E^{-}-T_{Q,j}$ so (for $E^{-}\in A_J$)
take derivatives in $Q(E^{-}-T_{Q,j})-
Q(E^{-})=T_{E,j}$ (cf. (\ref{CA}) applied to $E$ and $Q$!)
to obtain $\partial_{Q,j}Q^{+}-\partial_{Q,j}Q^{-}=
(dQ/dE)$ (i.e. $\partial T_{E,j}/\partial T_{Q,j}=0=\partial_{Q,j}Q^{+}-
\partial_{Q,j}Q^{-}-d Q/d E$) and this
implies $\partial_{Q,j}d{\cal S}(E,T)=d\Omega_{Q,j}$.  As a
corollary one concludes that $E(p,T)$ and $Q(p,T)$ satisfy the Whitham
eequations $\partial_AE=\{E,\Omega_A\}$ and $\partial_AQ=\{Q,\Omega_A\}$
plus the classical string equation $\{Q,E\}=1$.
\\[3mm]\indent {\bf REMARK 5.1.}$\,\,$ We make here a few remarks about
differentials with a view toward clarifying some matters above.
We recall first (following \cite{bb}) 
that an abelian diferential is a meromorphic one form
given locally via $\omega=f(z)dz$ where $f$ 
is meromorphic in its domain.  Evidently $d\omega=0$ and one defines
a primitive via $\Omega(P)=\int_{P_0}^P\omega$ on any simply connected
domain.  This is called an abelian integral and one must be careful in
their definition over the whole surface.  Thus let $(A_j,B_j)$ be a
canonical homology basis and define $\hat{A}_j=\int_{A_j}d\Omega$ with 
$\hat{B}_j=\int_{B_j}d\Omega$.  
Given a cycle $\gamma=\sum_i(m_iB_i+n_iA_i)$
one has $\int_{\gamma}d\Omega=\sum_i(m_i\hat{B}_i+n_i\hat{A}_i)$.  Abelian
integrals of the third kind have logarithmic singularities and it
is necessary to supplement the above cyclic periods by additional
polar periods in order to specify precisely the multivaluedness.
Thus if $d\Omega$ has poles at points $P_j$ one writes $c_j=\int_
{\gamma_j}d\Omega,\,\,j=1,\cdots,n$, where $\gamma_j$ is a cycle
homologous to zero containing $P_j$.  Evidently $c_j=2\pi iRes(d\Omega,
P_j)$ and necessarily $\sum_{sp}c_j=0$ (where $sp\sim$ singular  
points).  One writes
$\Sigma^0$ for the surface obtained by removing all $A$ and $B$ cycles.
Letting $A_j^{+},\,\,A_j^{-}$ and $B_j^{+},\,\,B_j^{-}$ be the left
and right edges of the appropriate cuts, $\Sigma^0$ has boundary
$\partial\Sigma^0=\sum_1^g(A_j^{+}+B_j^{+}-A_j^{-}-B_j^{-})$.
Any abelian integral $\Omega(P)$ of the first or second kind (i.e.
holomorphic or meromorphic with residues equal to zero at all singular
points) is single valued on $\Sigma^0$.  At the boundary it is required
that
\be
\left.\Omega(P)\right|_{A_j^{+}}-\left.\Omega(P)\right|_{A_j^{-}}=
-\hat{B}_j;\,\,\left.\Omega(P)\right|_{B_j^{+}}-\left.\Omega(P)\right|_
{B_j^{-}}=\hat{A}_j
\label{CA}
\ee
This allows continuation of $\Omega$ as a single valued function on
the universal cover of $\Sigma$.  To determine a single valued branch
of a third kind abelian integral one must draw additional cuts $\Gamma_j$
between $P_0$ and $P_j$, where $P_j$ are the singular points.
Then the polar period formulas are supplemented with 
\be
\left.\Omega(P)\right|_{\Gamma_j^{+}}-\left.\Omega(P)\right|_{\Gamma_j^{-}}
=c_j
\label{CB}
\ee
The Riemann bilinear relations yield
$$
(i)\,\,\sum_1^g(\hat{A}'_k\hat{B}_k-\hat{A}_k\hat{B}'_k)=0;
\,\,(ii)\,\,\sum_1^g(\hat{A}'_k\hat{B}_k-\hat{A}_k\hat{B}'_k)=
\frac{2\pi i}{(n-1)!}\frac{d^{n-1}}{dz^{n-1}}
\left.\Omega'(z)\right|_{z=z_0};$$
\be
(iii)\,\,\sum_1^g(\hat{A}'_k\hat{B}_k-\hat{A}_k\hat{B}'_k)=
2\pi i\sum_1^nc_j\Omega'(P_j)\equiv
2\pi i\sum_1^nc_j\int_{P_0}^{P_j}d\Omega'
\label{CC}
\ee
where in (i) $\Omega,\,\,\Omega'$ are integrals of the first kind, in
(ii) $\Omega'$ is of first kind and $\Omega$ is of the second kind with
a single pole at $P_0$ with local parameter chosen so that
$d\Omega=[(z-z_0)^{-n} +O(1)]dz,\,\,n>1$, and in (iii) $\Omega'$ is of
first kind and $\Omega$ is of third kind with no more than logarithmic
singularities at $P_j,\,\,j=1,\cdots,n$ and $c_j=Res(d\Omega,P_j)$
(the integration contours in (iii) are chosen to be distinct from
the basic $A$ or $B$ cycles).  As a consequence one knows that any
abelian differential of the second or third kind with zero a-periods
(b-periods) or with all purely imaginary (purely real) cyclic periods
is uniquely defined by its principal parts at the singular points.
The basic holomorphic differentials $d\omega_j$ are normalized by the
condition $\int_{A_j}d\omega_k=\delta_{jk}$ and for hyperelliptic
situations one can write
\be
d\omega_j=\sum_1^gc_{jk}\frac{\lambda^{g-k}d\lambda}{\mu}\,\,\,(1\leq j\leq g);
\,\,c_{jk}=2\pi i(A^{-1})_{jk};\,\,A_{jk}=\int_{A_k}\frac{\lambda^{g-j}d
\lambda}{\mu}
\label{CD}
\ee
Here the matrix $B_{jk}=\int_{B_j}d\omega_k$ is the period matrix
and due
to the placements of $2\pi i$ in the normalization one has 
$\Re B<0$.  
\\[3mm]\indent
{\bf REMARK 5.2.}$\,\,$ In summary, for dKP, ${\cal S}
=\sum_1^{\infty}{\cal B}_n
T_n$ is consistent with $\partial_n{\cal S}={\cal B}_n$ and $\partial_{\lambda}
{\cal S}={\cal M}$. 
Evidently $\partial_n{\cal S}={\cal B}_n$
and for completeness we show that $\partial_{\lambda}S={\cal M}$ (cf. 
\cite{cr}).  Thus from
the equations before (\ref{YEEE})
\be
{\cal S}=\sum_1^{\infty}T_n\lambda^n+\sum_1^{\infty}T_n\sum_1^{\infty}
\partial_nS_{j+1}\lambda^{-j}
\label{NL}
\ee
and one recalls that $2T=\sum_1^{\infty}T_n\partial_nF$.  Hence
\be
\partial_{\lambda}{\cal S}=\sum_1^{\infty}nT_n\lambda^{n-1}-\sum_1^{\infty}
T_n\sum_1^{\infty}j\partial_nS_{j+1}\lambda^{-j-1}=
\label{NM}
\ee
$$=\sum_1^{\infty}nT_n\lambda^{n-1}+\sum_1^{\infty}\sum_1^{\infty}T_nF_{jn}
\lambda^{-j-1}$$
But the last series is $\sum_1^{\infty}\lambda^{-j-1}[\partial_j\sum_{n=1}^
{\infty}T_nF_n-F_j]=\sum_1^{\infty}F_j\lambda^{-j-1}$ which
implies that $\partial_
{\lambda}{\cal S}={\cal M}$.  
For Riemann surfaces with $d{\cal S}$ as in (\ref{JK}) we can 
write $\partial_Ad{\cal S}
=d\Omega_A$ with $\oint_{A_i}d{\cal S}=a_i$ so that this is commensurate with
(\ref{JB}) for example where $T_{E,i}=T_{Q,i}=0$.  Note here $\tau_{B_i,E}=
\oint_{B_i}dE\sim-T_{Q,i}$ and $\tau_{B_i,Q}=\oint_{B_i}dQ\sim T_{E,i}$
and since these $\tau$'s are not used much in \cite{kt} they should
probably be renamed to correspond to their natural function here (i.e.
interchange the subscripts $E$ and $Q$).  In any event it appears that
dual variables $a_i^D$ to the action like ``times" $a_i$ could be defined
in Whitham theory via
\be
a_j^D=\sum_1^{\infty}T_n\oint_{B_j}d\Omega_n+\sum_1^ga_iB_{ij}+
\sum_1^gT_A\oint_{B_j}d\Omega_A
\label{JL}
\ee

\subsection{Symplectic geometry}

We sketch here the construction of \cite{kt} 
for the symplectic form $\omega_{{\cal M}}$ following \cite{co}.  Thus
go back to ${\cal M}_g(n,m)=\{\Sigma_g,P_1\sim\infty,[k]_n,[k]_m,E,Q\}$
defined at the beginning of Section 5.2, with times defined via (\ref{GC}).
Define ${\cal D},\,\,{\cal N},\,\,{\cal M},$ and ${\cal N}^g$ as before
and work on ${\cal N}$. We recall that
although $E$ and $Q$ are multivalued on
the universal fibration their differentials are well defined on ${\cal N}$.
$E$ and $Q$ are well defined near $P_1$ and their
analytic continuations by different paths can only change by multiples
of their residues or periods along closed cycles.  On a leaf of the
foliation the ambiguities remain constant and disappear upon differentiation.
Hence one has differentials $\delta E$ and $\delta Q$ on the fibrations
which reduce to $dE$ and $dQ$ acting on vectors tangent to the fiber. 
One can trivialize the fibration ${\cal N}$ with the variables
$a_1,\cdots,a_g$ along the leaf ${\cal M}$ and e.g. $E$ along the
fiber.  Then $dQ$ coincides with $(dQ/dE)dE$ where $dE\sim\delta E$ 
is viewed as one
of the elements of the basis of one forms for ${\cal N}$ 
and the full differential for $Q$ is $\delta Q=dQ+\sum_1^g(\partial Q/
\partial a_i)da_i\equiv dQ+\delta^EQ$.
\\[3mm]\indent
Now if one considers the full differential $\delta(QdE)$ on ${\cal N}$
it is readily seen that it is well defined despite the multivaluedness
of $Q$.  In fact the partial derivatives $\partial_{a_i}(QdE)$ along
the base ${\cal M}$ are holomorphic since the singular parts of the 
differentials as well as the ambiguities are all fixed.  In particular
$(\partial/\partial a_i)(QdE)=d\omega_i$
where $d\omega_i$ is a basis of normalized holomorphic differentials
$\oint_{A_i}d\omega_j=\delta_{ij}$ with $\oint_{B_i}d\omega_j=b_{ij}$.
To see this note that 
it is implicit in (\ref{GC}) since by definition of the $a_i,\,\,
(\partial a_i/\partial a_j)=\delta_{ij}=\oint_{A_i}(\partial (QdE)/
\partial a_j)$ which implies $(\partial (QdE)/\partial a_j)=d\omega_j$.
Formally then one defines $\omega_{{\cal M}}$ on ${\cal N}^g$ via
\be
\omega_{{\cal M}}=\delta\left(\sum_1^gQ(\gamma_i)dE(\gamma_i)\right)=
\sum_1^g\delta Q(\gamma_i)\wedge dE(\gamma_i)=\sum_1^gda_i\wedge d\omega_i
\label{GF}
\ee
The first expression seems formally reasonable on ${\cal N}^g$ and the last
appears to be a calculation of the form
$\omega_{{\cal M}}=\delta(QdE)=\sum_1^gda_i\wedge [\partial(QdE)/\partial
a_i]=\sum_1^gda_i\wedge d\omega_i$.
Now go to KP for illustration and background.  One can work with 
a (nonspecial) divisor
$(\gamma_1,\cdots,\gamma_g)$ giving rise to quasiperiodic
functions of $t=(t_n)\,\,1\leq n<\infty,$ of the form $u_{i,n}
\,\,1\leq i\leq n-2,\,\,2\leq n<\infty$, which arise as solutions of an
integrable hierarchy.  The BA function is defined as a meromorphic function
away from $P$ with simple poles at the $\gamma_i\,\,(1\leq i\leq g)$
and an essential singularity at $P$ of the form
$\psi(t,z)=exp\left(\sum_1^{\infty}t_nz^{-n}\right)
\left(1+\sum_1^{\infty}\xi_i(t)z^i\right)$.
There is a Lax operator (in general one for each puncture)
$L_n=\partial^n+
\sum_0^{n-2}u_{i,n}\partial^i$
with $\left(\partial/\partial t_n-L_n\right)
\psi(t,z)=0\,\,(\partial=\partial_x,\,\,x=t_1)$.
Thus there is a map $\{\Sigma,P_,z,
\gamma_1,\cdots,\gamma_g\}\to \{u_{i,n}(t)\}$.  
An explicit form for the BA function for KP is given in \cite{kt}
(recall $\Re\oint_Cd\Omega_n=0$ for any cycle
$C$ here and cf. \cite{bb} for various forms of BA functions - thus
the form of $\psi$ in \cite{kt} appears different from (\ref{psi})
but it must be equivalent).
Similarly the dual BA function $\psi^*$ is 
defined as before (and denoted by $\psi^{\dagger}$ in \cite{kt}).
We note that for the Toda lattice one takes $N=2$ punctures
and a common theme here to some other work is 
$d(\lambda_{SW})\sim\omega_{{\cal M}}$ in a suitable sense (cf. Section 5.2).
\\[3mm]\indent
An element
in ${\cal N}^g(n,1)$ gives rise to a datum in a space 
$\hat{{\cal N}}^g =\{u_{in}(t)|_0^{n-2}\}$ via a mapping
$\Xi:\,\,(\Sigma,P,[z]_n,E,Q,\gamma_1,\cdots,\gamma_g)\to (\Sigma,
P,z,\gamma_1,\cdots,\gamma_g)\to \{\left.u_{i,n}(t)\right|_{i=0}^{n-2}\}$.
Take a real leaf ${\cal M}$ (i.e. $\Re\oint_CdE=\Re\oint_CdQ=0$
for all cycles $C$ on $\Gamma$ and note this is automatic for $dQ\sim dp
\sim d\Omega_1$ and $dE\sim d\Omega_n$ with real normalization).
We also take $t_1\sim x$.  One wants now
to express $\omega_{{\cal M}}$ in terms of forms on the space of functions
$\{u_{i,n}(t)\}$.  First the $u_{i,n}(t)$ can be written in terms of the 
asymptotic BA coefficients
$\xi_i$ and one knows that
$(\partial_x\psi/\psi)=z^{-1}+\sum_1^{\infty}h_sz^s$ and $
(\partial_x\psi^*/\psi^*)=-z^{-1}+\sum_1^{\infty}h^*_sz^s$
(cf. \cite{cd} and Lemma 3.1, where e.g.
$\partial = L+\sum_1^{\infty}\sigma_j^1L^{-j}$  which 
implies e.g. $(\partial\psi/\psi)=\lambda+\sum_1^{\infty}\sigma_j^1
\lambda^{-j}$ for $\lambda\sim z^{-1})$.  In any event the first $n-1$
coefficients $h_s=\sigma_s^1$ or $h^*_s$ are differential polynomials
in the $u_{i,n}$ (initial data $\left.\xi_s(t)\right|_{x=0}=\phi_s(t_2,\cdots)$
determine $\xi_s$ for $s\leq n-1$).  Writing $H_s=<h_s>$
one gets now $p=z^{-1}+\sum_1^{\infty}
H_sz^s$ (recall also from Lemma 3.1 that $<s_{n+1}>=-n<\sigma_n^1>=-n
H_n$).  
We note that one uses $<\,\,\,>_x$ and $<\,\,\,>_{xy}$
averaging at various places in \cite{kt} but generically this should
correspond to ergodic averaging.  
\\[3mm]\indent
A result in \cite{kt} now asserts that
for ${\cal N}^g$ the Jacobian
bundle over a real leaf ${\cal M}$ of the moduli space ${\cal M}_g(n,1)$
the symplectic form $\omega_{{\cal M}}$ can be written as
\be
\omega_{{\cal M}}=Res_P\frac{<\delta\psi^*\wedge
\delta L_n\psi>}{<\psi^*\psi>}dp
=\sum_1^{n-1}<\delta J_s\wedge\int^x\delta^*h^*_{n-s}>
\label{GK}
\ee
Here $h_s,\,\,h^*_s$ are differential polynomials as above, the
differential forms $\delta h_s$ and $\delta^*h^*_s$ are defined via
$\delta h_s=\sum_{i=0}^{n-2}\delta u_{i,n}\sum_{\ell}(-\partial)^{\ell}
(\partial h_s/\partial u_{i,n}^{(\ell)})$ and, in a standard manner,  
$\delta^*h^*_s=\sum_{i=0}^{n-2}\delta u_{i,n}^{(\ell)}\sum
(\partial h^*_s/\partial u_{i,n}^{(\ell)})$, while $\delta J_s$ is
a linear combination of the $\delta u_{in},\,\,(0\leq i\leq n-2)$. Another
(essentially more complicated)
form is also given in \cite{kt} for the last term in (\ref{GK}) 
and we will give some evidence for its validity without
actually proving it (see below).  In any event
this is a fascinating result but the proof in \cite{kt} requires some
embellishment which we extract here from \cite{co} with some sign changes.  
First one must come to terms with an expression
\be
\delta E=\delta p\left(\frac{dE}{dp}\right)-
\left(\frac{<\psi^*\delta L_n\psi>}{<\psi^*\psi>}\right)\equiv
\label{GII}
\ee
$$\equiv \delta p=\left(\frac{dp}{dE}\right)\delta E
+\left(\frac{<\psi^*\delta L_n
\psi>}{\psi^*\psi>}\right)\left(\frac{dp}{dE}\right)$$
where $\delta L_n=\sum_0^{n-2}\delta u_{i,n}\partial^i$
(see below) and then it will follow from (\ref{GF}) (with $\delta p\sim
\delta Q$ and $\delta E\sim dE$) that 
\be
\omega_{{\cal M}}=\sum_1^g\left(\frac{<\psi^*
\delta L_n\psi>}{<\psi^*\psi>}\right)(\gamma_s)\wedge  dp(\gamma_s)
\label{GGG}
\ee
The formula (\ref{GII}) is asserted to come from \cite{kc}
but to see this one has to interpret $\delta u_{i,n}$ as arising from
$\epsilon\partial_XU_{i,n}(X,T)$ in the first order term.  Indeed we
can look at (\ref{BG}) with the last two terms absent on the leaves of
our foliation and written generically for $L_n\sim \Omega_n$ as
$(\bullet\spadesuit)\,\,<\psi^*(L_n^1\partial_XL_k-L_k^1\partial_XL_n)
\psi>=\partial_X\Omega_k<\psi^*L_n^1\psi>-\partial_X\Omega_n
<\psi^*L_k^1\psi>$.  Then for $L_n=L_3\sim\Omega_3$ and $L_k=L_1\sim
p$ one has $L_1=\partial,\,\,\delta L_1\sim\partial_XL_1=0,\,\,L_1^1=-1$
and $-<\psi^*\partial_XL_3\psi>=\partial_Xp<\psi^*L_3^1\psi>+\partial_X
\Omega_3<\psi^*\psi>$.  Written in terms of $\delta p\sim\epsilon
\partial_Xp,\,\,\delta L_3\sim\epsilon\partial_XL_3$, etc. one obtains
$\delta\Omega_3=-\delta p(<\psi^*L_3^1\psi>/<\psi^*\psi>)-(<\psi^*
\delta L_3\psi>/<\psi^*\psi>)$ while from Theorem 3.2 (with $L_3^1\sim
A^1$) we have $(<\psi^*L_3^1\psi>/<\psi^*\psi>)=-(d\Omega_3/dp)$, 
or generically $(d\Omega_n/dp)=-(<\psi^*L_n^1\psi>/
<\psi^*\psi>)$, and 
(\ref{GII}) follows.  
The formula $(\bullet\spadesuit)$ can also be used
for a more general theorem in \cite{kt}.  Thus (\ref{GGG}) holds and
writing $\gamma_s(t)$ for the zeros of $\psi$ away from $P$ (corresponding
by Riemann-Roch
to the fixed poles $\gamma_s$ for $t=0$) one can pick $t_1,\cdots,t_g$ 
(generically) as times for which the flows $\gamma_s(t)$ are independent
and use them as coordinates on $S^g(\Sigma)$.  These can be transferred to
the system of coordinates $f(\gamma_1),\cdots,f(\gamma_g))$ for $f$ an
Abelian integral on $\Sigma$ via 
$(\partial/\partial t_i)f(\gamma(t))=Res_{\gamma(t)}
[(\partial/\partial t_i)\psi(t,z)/\psi(t,z)]df$.  In the present
case, writing $(\delta_t\psi/\psi)=\sum_1^g(\partial_j\psi/\psi)dt_j$ and 
using $\sum Res_{\gamma_s}=-Res_P$ one obtains 
$\omega_{{\cal M}}=Res_P\left[(<\psi^*\delta L_n\psi>/
<\psi^*\psi>)\wedge
\left.(\delta_t\psi/\psi)\right|_{t=0}\right]dp$.
It is not unnatural to see here an apparent
change in the number of parameters. 
\\[3mm]\indent 
Now note that $\psi^*\delta L_n\psi=
\psi^*\sum_0^{n-2}\delta u_{i,n}\partial^i\psi
=\sum_0^{n-2}\delta u_{i,n}\psi^*\partial^i\psi$ and we recall
$\partial\psi/\psi=z^{-1}+\sum_1^{\infty}h_sz^s$ from which
$(\partial^j\psi/\psi)=z^{-j}[1+\sum_1^{\infty}c_{jp}z^{p+1}]\,\,
(c_{jp}=c_{jp}(h_s,\partial^ih_s))$.  Then
since $dp/<\psi^*\psi>=[-z^{-2}+O(1)]dz$ and $\psi^*\psi=1+\sum_1^{\infty}
s_nz^n$ for small $z$, we have
\be
\left(\frac{\psi^*\delta L_n\psi}{<\psi^*\psi>}\right)dp\sim
\frac{\sum_0^{n-2}\delta
u_{in}\psi^*\partial^i\psi dp}{<\psi^*\psi>}\sim
\label{GHH}
\ee
$$\sim -\sum_1^{n-2}\delta u_{in}z^{-i}\left(1+\sum_1^{\infty}
c_{ip}z^{p+1}\right)(z^{-2}+o(1))(1+\sum_1^{\infty}s_nz^n)dz\sim
\sum_1^{\infty}\delta J_sz^{-n-1+s}dz$$
where $\delta J_s$ is by construction a linear combination of the 
$\delta u_{in},\,\,(0\leq i\leq n-2)$.  Averaging now in (\ref{GHH}) and 
using (\ref{GII}) one obtains $<\delta J_s>=0$ for $s=1,\cdots,n-1$.
In addition it is claimed in \cite{kt} that $\delta H_s=0$ for
$s=1,\cdots,n-1$, since the mean values $H_s=<h_s>$ are fixed along
${\cal M}$ (this is not entirely clear).  Hence 
$\omega_{{\cal M}}=\sum_1^g<\delta J_{j+n}>\wedge\,\, dt_j$.
Further from \cite{cd,ci} one has 
$(\partial_j\psi^*/\psi^*)=-\lambda^j-\sum_1^{\infty}\sigma_s^j\lambda^
{-s}=-z^{-j}+\sum_1^{\infty}h^*_{s,j}z^s$.  This leads to
\be
\frac{\partial_j\psi^*(\delta L_n\psi)}{<\psi^*\psi>}dp=\frac{\partial_j
\psi^*}{\psi^*}\times\frac{\psi^*(\delta L_n\psi)}{<\psi^*\psi>}dp=
\label{GZZ}
\ee
$$=\left(-z^{-j}+\sum_1^{\infty}h^*_{p,j}z^p\right)\times\left(-\sum_1^{\infty}
\delta J_s(t)z^{-n-1+s}\right)dz$$
which implies
\be
\delta J_{j+n}(t)=\sum_1^{n-1}\delta J_sh^*_{n-s,j}
\label{HAA}
\ee
Hence
$\omega_{{\cal M}}=Res_P(<\delta_t\psi^*\wedge \delta L_n\psi>/
<\psi^*\psi>)dp$ holds 
and an argument is suggested in \cite{kt}
to extend $\delta_t$ to a full $\delta$ so that the first formula
in (\ref{GK}) is proved (we have made some adjustments in minus signs
from \cite{co,kt}). 
For the second formula in (\ref{GK}) one looks
at the relation $\delta log\psi^*=
\delta\left(c(t_i,\,i\geq 2)+\int_{x_0}^x\partial_xlog
\psi^*\right)=
\delta\sum_1^{\infty}\left(c_s(t_i,\,i\geq 2)+\int_{x_0}^xh_s^*z^sdx
\right)$.  This implies
\be 
Res_P\left(\frac{<(\delta log\psi^*)\wedge (\psi^*\delta L_n\psi)>)}
{<\psi^*\psi>}\right)dp=
\omega_{{\cal M}}=\sum_1^{n-1}<\delta J_s\wedge\int^x\delta^*
h^*_{n-s}dx>
\label{GKK}
\ee
which is the second identity in (\ref{GK}).  In \cite{kt} one observes that
by definitions $\delta J_s$ does not contain variations of derivatives
of $u_i$ and then claims that
$\delta J_s=-n(\partial h_s/\partial u)\delta u\sim -n\delta h_s$.
This seems unclear but we will write out some relevant formulas below.
Finally we note that there
is a little interplay between $\delta\sim\epsilon\partial_X$ and $X=
\epsilon x$ in the last two lines which we do not elaborate upon here
(cf. \cite{cr} for further details of calculations, etc.).  
The examples in \cite{kt} have some
curious features but in any event one should keep in mind the conditions
$\delta H_s=0$ when selecting functions.  One notes that 
of course the symplectic
structure involving $\delta J_s,\,\,\delta^* h_s^*$, etc. 
(i.e. the $\delta u_{in}$) does not
include the time dynamics so KdV $\sim L_2=\partial^2+u$ for example
and Bousinesq $\sim L_3=\partial^3+u\partial+v$.  Adaptions and 
applications of the above fibration framework to $SU(n),\,\,N=2$ susy
YM theory are also given in \cite{kt}.
\\[3mm]\indent
In order to compare the $\delta J_s$ and $\delta h_s$ we make a few
calculations here.  Thus $\delta J_s$ is given via (\ref{GHH}) as a linear
combination of the $\delta u_{in}$ with complicated coefficients 
depending on $h_s,\,\partial^ih_s$ and the coefficients $s_n$ in $\psi^*
\psi$ (note the $s_n$ can be expressed via $\xi_i$ and $\xi^*_i$ while
the term $dp/<\psi^*\psi>$ can be written $(z^{-2} +\sum_0^{\infty}
p_mz^m)dz$ where the $p_m$ are independent of the $u_{in}$).  We recall here
(cf. \cite{ca,cd,mw}) for $\psi=exp(\xi)[1+\sum_1^{\infty}\xi_jz^j]$ and
$\psi^*=exp(-\xi)[1+\sum_1^{\infty}\xi^*_iz^i]$, with $L=\partial +
\sum_1^{\infty}u_{i+1}\partial^{-i}\,\,(u_2\sim u)$ one has $u_2=-\xi_1',\,\,
u_3=-\xi'_2+\xi_1\xi'_1,$ etc.  Further from $\partial\psi=z^{-1}\psi
+exp(\xi)\sum\xi'_iz^i$ and $(\partial\psi/\psi)=z^{-1}+\sum h_sz^s$ we get
\be
(1+\sum_1^{\infty}\xi_jz^j)(\sum_1^{\infty}h_sz^s)=\sum_1^{\infty}\xi'_kz^k
\Rightarrow\,\,\xi'_m=h_m+\sum_1^{n-1}\xi_jh_{n-j}
\label{GXX}
\ee
so e.g. $h_1=\xi'_1,\,\,\xi'_2=\xi_1h_1+h_2,\,\,\xi'_3=h_3+h_2\xi_1+\xi_2h_1,$
etc.  Hence $h_1=-u_2,\,\,h_2=\xi'_2-\xi_1h_1=\xi_1\xi'_1-u_3+\xi_1u_2=-u_3,$
etc.  In the present situation one is interested in $L^n_{+}\sim L_n$ so we 
note that $L^2_{+}=\partial^2+2u_2,\,\,L^3_{+}=\partial^3+3u_2\partial+
(6u'_2+3u_3),$ etc. (we recall also $L_n=\partial^n+\sum_0^{n-2}u_{in}
\partial^i$).  Thus for $n=3$ we have $u_{03}=3u_3+6u'_2,\,\,u_{13}=
3u_2,$ etc.  
Calculations based on this seem to agree with the formulas
in \cite{kt}, namely, for fixed $n$
\be
h_1=-\frac{u_{n-2,n}}{n};\,\,h_2=\frac{n-1}{n}u'_{n-2,n}-\frac{u_{n-3,n}}{n};
\,\,\cdots
\label{GYY}
\ee
(formulas for $h^*_i$ are also given in \cite{kt}). 
Thus for $n=3$ one has $h_1=-(1/3)u_{13}=-u_2$ and $h_2=(2/3)u'_{13}-
(1/3)u_{03}=-u_3+2u'_2-u'_2=-u_3$, etc. 
As for the $Q_s$ we
write e.g.
\be
\frac{\partial^2\psi}{\psi}=[z^{-1}+\sum_1^{\infty}h_sz^s]^2 +
\sum_1^{\infty}h'_sz^s=
\label{GGXX}
\ee
$$=z^{-2}[1+2\sum h_sz^{s+1}+2\sum\sum_{s\not= k}h_sh_kz^{s+k+2}+
\sum h_s^2z^{2s+2}+\sum h'_sz^{s+2}]$$
Then writing this as $z^{-2}[1+\sum_1^{\infty}c_{2p}z^{p+1}]$ one has
$(\bullet\clubsuit)\,\,c_{21}=2h_1,\,\,c_{22}=2h_2+h'_1,\,\,
c_{23}=2h_3+h_1^2+h'_2,\cdots$.  Now look at (\ref{GHH}) for $n=3$ so we
have to consider $\delta u_{03}\psi$ and $\delta u_{13}\partial\psi$, 
yielding for $\Xi=\psi^*\delta L_3\psi$ the formula
\be
\Xi=[\delta u_{03}+\delta u_{13}(z^{-1}+\sum h_sz^s)](1+\sum s_mz^m)
\label{GGYY}
\ee
Putting in the $(dp/<\psi^*\psi>)$ term
we obtain for $n=3$
\be
(\ref{GHH})=\sum_1^{\infty}\delta J_sz^{s-4}dz=
[\delta u_{13}z^{-3}+\delta u_{03}z^{-2}]\times
\label{GGZZ}
\ee
$$\times [1+\sum_1^{\infty}s_mz^m][1+\sum_1^{\infty}h_sz^{s+1}][1
+\sum_0^{\infty}p_kz^{k+2}]dz=$$
$$=[\delta u_{13}z^{-3}+\delta u_{03}z^{-2}]
[1+s_1z+(s_2+h_1+p_0)z^2+\cdots]dz$$
Consequently $\delta J_1=\delta u_{13}=3\delta u_2=-3\delta h_1,
\,\,\delta J_2=
s_1\delta u_{13}+\delta u_{03}$, etc.  Note $s_1=\xi_1+\xi^*_1=0$ so
$\delta J_2=\delta u_{03}=\delta(3u_3+6u'_2)=-3\delta h_2+6\delta u'_2$
(note also $h_2\sim -3u_{03}$ mod $u'_2$).
We see that in some ``generic" manner the relation $\delta J_n\sim -n\delta
h_n$ seems to be holding (modulo terms involving derivatives of the
$u_{in}$).  Hence provisionally one could say that there is some evidence
for a formula of the form
\be
\omega_{{\cal M}}\sim-n\sum_1^{n-1}<\delta h_s\wedge\int^x\delta^*
h^*_{n-s}>
\label{GGGG}
\ee
as in \cite{kt}, where $\delta h_s$ only deals with the appropriate
potentials and not their derivatives.
However the expression (\ref{GK}) in terms of $\delta J_s$ is simpler and
more direct so we treat this as the basic such formula.

\section{REMARKS ON BA FUNCTIONS}
\renewcommand{\theequation}{6.\arabic{equation}}\setcounter{equation}{0}

We go here to \cite{ne} for a construction of BA functions with certain
$\alpha_i$ variables inserted explicitly which connect to the $a_i$ ``times"
of SW theory via $i\epsilon\alpha_j=a_j$.  This leads to expressions for
${\cal S}$ and $F$ (action and prepotential) in two different forms from
which some comparisons can be made, of use in various directions (cf.
\cite{cp,ib,ne}).  The construction in \cite{ne} is for the Toda theory
(with two punctures on the Riemann surface) but, in keeping with the spirit
of this paper, we will display matters for the KP situation only.  The idea
is to insert new times $\alpha_i$ in the BA function (\ref{psi}) 
(recall $B_{ij}=\oint_{B_j}d\omega_i,\,\,U=(U_j)=
(1/2\pi i)\oint_{B_j}d\Omega^1$,
etc.) via for example
\be
\psi=exp\left[\int_{P_0}^Pxd\Omega^1+yd\Omega^2+td\Omega^3+i\sum_1^g
\alpha_id\omega_i\right]\times
\label{alpha}
\ee
$$\times\,\,\frac{\Theta\left({\cal A}(P)+xU+yV+tW+i\sum_1^g
\alpha_i(B_{ij})+z_0\right)}{\Theta({\cal A}(P)+z_0)}$$
This is an alternate version of the BA function in \cite{ne}, modulo
$\bar{t_n}=0$ (cf. \cite{bb}) for relations)
or one can adapt the formulas of \cite{ne} to write (recall ${\cal A}(P)=
(\int_{P_0}^Pd\omega_j)$ and we set for convenience $\Omega_{jk}=\int_{B_k}
d\Omega_j$ where $d\Omega_j\sim d\Omega^j$)
\be
\psi=exp\left[\sum_1^{\infty}t_j\left(\int_{P_0}^Pd\Omega^j+\Omega^j(P_0)
\right)+i\sum_1^g
\alpha_j\left(\int_{P_0}^Pd\omega_j+\omega_j(P_0)\right)\right]\times
\label{omega}
\ee
$$\times\frac{\Theta\left({\cal A}(P)+\sum_1^{\infty}(t_j/
2\pi i)(\Omega_{jk})
+i\sum_1^g\alpha_j(B_{jk})+z_0\right)\Theta({\cal A}(P_{\infty})
+z_0)}{\Theta\left({\cal A}(P_{\infty})+\sum_1^{\infty}
(t_j/2\pi i)(\Omega_{jk})
+i\sum_1^g\alpha_j(B_{jk})+z_0\right)\Theta({\cal A}(P)+z_0)}$$
(note one can write here also $\int^Pd\Omega_j\sim\int_{P_0}^Pd\Omega_j
+\Omega_j(P_0)$) and ${\cal A}(P)=(\int_{P_{\infty}}^Pd\omega_j)+{\cal A}
(P_{\infty})$, 
where $\Omega_j\sim\lambda^j-\sum_1^{\infty}
(q_{mj}/m)z^{m-1}$).
We have attached a factor of $(1/2\pi i)$ to the $(\Omega_{jk})$ in
order to utilize (\ref{EWW}) below.
This kind of formula (\ref{omega})
provides a normalization as $z\to 0\,\,(\lambda\to\infty$) since, for 
any $P_0$, the theta function quotient tends to one.
Next one constructs in an
ad hoc manner the algebraic form of $\psi$ near $\infty$ via vertex 
operators as ($\lambda\sim z^{-1}$)
\be
\psi=exp\left(\sum_1^{\infty}t_i\lambda^i\right)\times\frac{\tau
(t-[\lambda^{-1}],\alpha)}{\tau(t,\alpha)}
\label{beta}
\ee
(here $[\lambda^{-1}]\sim(1/n\lambda^n)$).  Note that the $\alpha_i
\int^Pd\omega_i$ terms do not contribute to the essential singularity
at $\infty$.  
\\[3mm]\indent
Now the general action and prepotential theory associated with (\ref
{alpha}) (cf. \cite{bd,co,cr,ib,kz,ne}) leads to
\be
d{\cal S}=\sum_1^ga_jd\omega_j+\sum_1^{\infty}T_nd\Omega_n;\,\,
\frac{\partial d{\cal S}}{\partial a_j}=d\omega_j;\,\,\frac{\partial d{\cal S}}
{\partial T_n}=d\Omega_n
\label{gamma}
\ee
If we consider functions $F(a,T)$ related to $d{\cal S}$ via
\be
\frac{\partial F}{\partial a_j}=\frac{1}{2\pi i}\oint_{B_j}d{\cal S};\,\,
\partial_nF=-Res_{\infty}z^{-n}d{\cal S}
\label{delta}
\ee
then, given the standard 
class of solutions of the Whitham hierarchy satisfying
\be
2F=\sum_1^ga_j\frac{\partial F}{\partial a_j}+\sum_1^{\infty}T_n\frac
{\partial F}{\partial T_n}
\label{epsilon}
\ee
there results
\be
2F=\sum_1^g\frac{a_j}{2\pi i}\oint_{B_j}d{\cal S}-\sum_1^{\infty}
T_nRes_{\infty}z^{-n}d{\cal S}
\label{mumu}
\ee
(connections to \cite{kj} for example will be spelled out below).
Writing now, in the notation of \cite{ne},
$d\omega_j=-\sum_1^{\infty}\sigma_{jm}z^{m-1}dz$ with $d\Omega_n=
[-nz^{-n-1}-\sum_1^{\infty}q_{mn}z^{m-1}]dz$, and using (\ref{gamma}),
(\ref{mu}) can be expanded to give 
\be
2F=\frac{1}{2\pi i}\sum_{j,k=1}^gB_{jk}a_ja_k+2\sum_1^ga_j\sum_1^{\infty}
\sigma_{jk}T_k+\sum_{k,l=1}^{\infty}q_{kl}T_kT_l
\label{nu}
\ee
\indent
Thus the expression (\ref{nu}) comes from the Riemann surface theory, 
without explicit reference to the BA function, and we
consider now (\ref{omega}) and (\ref{beta}) 
to which ideas of dKP can be applied to 
introduce the slow variables $T_k$.  This means that we will
be able to introduce
slow variables in two different ways and the resulting comparisons will show
an equivalence of procedures.  In practice this will enable one to treat
$\epsilon$ on the same footing in the Whitham theory and in the dispersionless
theory, and subsequent analysis will verify a construction in \cite{cp}.
Thus from (\ref{omega}) and (\ref{beta}) one obtains 
an expression for $\tau$ of the form
($t_1=x,\,\,t_2=y,\,\,t_3=t,\,\cdots$)
\be
\tau(t,\alpha)=exp[\hat{F}(\alpha,t)]\Theta\left(
{\cal A}(P_{\infty})+\sum_1^{\infty}(t_j/2\pi i)(\Omega_{jk})
+i\sum_1^g\alpha_j(B_{jk})+z_0\right)
\label{sigma}
\ee
where $k=1,\cdots,g$ and
\be
\hat{F}(\alpha,t)=\frac{1}{2}\sum_{k,l=1}^{\infty}q_{kl}t_kt_l-\frac
{1}{4\pi i}\sum_{j,k=1}^{\infty}B_{jk}\alpha_j\alpha_k+i\sum_1^g
\alpha_j\sum_1^{\infty}\sigma_{jk}t_k+\sum_1^{\infty}d_kt_k
\label{tau}
\ee
(see also \cite{ks} for a similar form - recall here ${\cal A}(P)=
(\int_{P_0}^Pd\omega_j)$ and $P_0\not= P_{\infty}$ is required - note
that we have inserted a factor of $1/2\pi i$ with the $(\Omega_{jk})$).
Putting in the slow variables $T_k=\epsilon t_k$ and $a_k=i\epsilon\alpha_k$
one will find
that the quadratic part of $\hat{F}$ 
in $T$ is exactly $F(a,T)/\epsilon^2$.
One recalls here that $\tau$ will have the form (\ref{YBB}) so that
$(F(a,T)/\epsilon^2)+(\sum d_kT_k/\epsilon)$ is a natural object
(modulo a multiplicative factor of lower order arising from the theta
function in (\ref{sigma}), which could be absorbed in an $exp(A/\epsilon)$
term). 
To say this another way, (\ref{YBB}) is the natural form of $\tau$ based
on (\ref{beta}) and it is associated with $\psi\sim exp[(1/\epsilon)
S+O(1)]$ as in (\ref{YBB}).  Given the Riemann surface background based
on (\ref{omega}), one is expressing the $O(1/\epsilon)$ term in
$\tau=exp[(1/\epsilon^2)F+O(1/\epsilon)]$ as $exp[(1/\epsilon)\sum d_kT_k]$
times a suitable asymptotic expansion of theta functions (for such
expansions see \cite{dc}, pp. 40-49).
In \cite{ne} one writes then from (\ref{sigma}) and (\ref{alpha})
respectively
\be
log\tau\left(\frac{T}{\epsilon},\frac{a}{i\epsilon}\right)=\epsilon^{-2}
\sum_0^{\infty}\epsilon^nF^{(n)}(T,a);
\label{zeta}
\ee
$$dlog\psi\left(p,\frac{T}{\epsilon},
\frac{a}{i\epsilon}\right)=\epsilon^{-1}\sum_0^{\infty}\epsilon^n
d{\cal S}^{(n)}(p,T,a)$$
where $d{\cal S}^{(0)}\sim d{\cal S}$ in (\ref{gamma}) and $F^{(0)}\sim F$
in (\ref{nu}).
\\[3mm]\indent
Let us spell out some of the details now.  Thus one notes in (\ref{omega})
that as $P\to P_{\infty}$ the theta function terms go to one and the
asymptotic behavior is 
mainly determined by the exponent.  Now consider identifying
(\ref{omega}) and (\ref{beta}) as $P\to P_{\infty}\,\,(P\to P_{\infty}\sim
z\to 0\sim \lambda\to\infty$) via ($\Theta_{asy}\sim$ 
theta quotient in (\ref{omega}))
\be
\frac{\tau_{-}}{\tau}e^{\sum_1^{\infty}t_j\lambda^j}\sim exp\left[
\sum_1^{\infty}t_j\left(\lambda^j-\sum_1^{\infty}q_{mj}\frac{z^m}{m}\right)
-i\sum_1^g\alpha_j\sum_1^{\infty}\sigma_{jm}
\frac{z^m}{m}\right]\Theta_{asy}
\label{pi}
\ee
(write $\omega_j(P_0)=\int_{P_{\infty}}^{P_0}d\omega_j$ and
note that $d(z^m)=mz^{m-1}dz=-m\lambda^{-m-1}
d\lambda$).  If we accept an ansatz (\ref{sigma}) - (\ref{tau}), 
then e.g. (\ref{pi}) can be written out as indicated below.  We write
$i\sum_1^g\alpha_j(B_{jk})+z_0=\hat{z}(\alpha)$ with 
$\hat{z}(\alpha)+\sum_1^{\infty}(t_j/2\pi i)(\Omega_{jk})=\tilde{z}(\alpha,t)$
and $\Delta\hat{F}=\hat{F}(\alpha,t-[z])-\hat{F}(\alpha,t)$.
Now look at $\Delta\hat{F}$ from (\ref{tau}) as
\be
\Delta\hat{F}=\frac{1}{2}\sum_{k,l=1}^{\infty}q_{kl}\left[\left(
t_k-\frac{z^k}{k}
\right)\left(t_l-\frac{z^l}{l}\right)-t_kt_l\right]-i\sum_{j=1}^g
\sum_{k+1}^{\infty}\alpha_j\sigma_{jk}\frac{z^k}{k}-\sum_1^{\infty}d_k
\frac{z^k}{k}=
\label{EA}
\ee
$$=-\frac{1}{2}\sum_{k,l=1}^{\infty}q_{kl}\left[\left(t_k\frac{z^l}{l}+
t_l\frac{z^k}{k}\right)-\frac{z^{k+l}}{kl}\right]-i\sum_{j=1}^g\sum_{k=1}^
{\infty}\alpha_j\sigma_{jk}
\frac{z^k}{k}-\sum_1^{\infty}d_k\frac{z^k}{k}$$
From (\ref{pi}) we want then to balance
\be
log\Theta_{asy}\sim
\Delta\hat{F}+\sum_1^{\infty}t_j\sum_1^{\infty}q_{mj}\frac{z^m}{m}+
i\sum_1^g\alpha_j\sum_1^{\infty}\sigma_{jm}\frac{z^m}{m}+
\label{EB}
\ee
$$+log\Theta\left({\cal A}(P_{\infty})-
\sum_1^{\infty}\frac{z^j}{j}\frac{(\Omega_
{jk})}{2\pi i}+\tilde{z}\right)-\log\Theta\left({\cal A}(P_{\infty})+
\tilde{z}\right)$$
Note also that ${\cal A}(P)-{\cal A}(P_{\infty})
\sim (\int_{P_{\infty}}^Pd\omega_k)\sim-\sum_1^{\infty}
\sigma_{km}(z^m/m)$.  
The balance demanded then involves at first order a balance of the explicit
$t_k$ terms with the remaining terms lumped together in $d_k$.
In particular this seems to require that 
$(\bullet\clubsuit\bullet)\,\,q_{kl}=q_{lk}$ 
(which is true since the sum of residues of a meromorphic
differential is zero - cf. (\ref{ES})) so that
$\sum q_{kl}t_k(z^l/l)=\sum t_k\sum q_{lk}(z^l/l)$.  Then the explicit
$t_k$ terms in $\Delta\hat{F}$ cancel the term $\sum t_j\sum q_{mj}(z^m/m)$ in
(\ref{EB}) and we can determine $d_k$ provided the theta function terms
have a suitable expansion.
\\[3mm]\indent
In fact one can look at the theta function balance by directly comparing
\be
A=\frac{\Theta\left({\cal A}(P_{\infty})-\sum\frac{z^j}{j}\frac
{(\Omega_{jk})}{2\pi i}+\tilde{z}\right)}{\Theta\left({\cal A}(P_{\infty})+
\tilde{z}\right)};
\label{EU}
\ee
$$B=\frac{\Theta({\cal A}(P)+\tilde{z})\Theta({\cal A}(P_{\infty})+z_0)}
{\Theta({\cal A}(P_{\infty})+\tilde{z})\Theta({\cal A}(P)+z_0)}$$
First we note some formulas from \cite{bb,cn,cq,sc}.  Using the
Riemann bilinear relations one can show that if $d\tilde{\Omega}_n=
(z^{-n}+O(z))dz$ then
\be
\oint_{B_j}d\tilde{\Omega}_n=\tilde{\Omega}_{nj}=-\frac{2\pi i}{(n-1)!}
\partial^n_z\omega_j|_{z=0}=-\frac{2\pi i}{(n-1)}\sigma_{j,n-1}
\label{EUU}
\ee
where $d\omega_j=f_jdz=-\sum_1^{\infty}\sigma_{jm}z^{m-1}dz=-\sum_0^{\infty}
\sigma_{j,p+1}z^pdz$ with $\omega_j\sim-\sum_1^{\infty}\sigma_{jm}(z^m/m)$.
Note also that $\partial^{n-1}f_j|_{z=0}=-\sigma_{jn}(n-1)!$ so
\be
\tilde{\Omega}_{n+1,j}=-\frac{2\pi i}{n}\sigma_{jn}=
\left.-\frac{2\pi i}{n!}\partial^{n-1}f_j\right|_{z=0}
\label{EVV}
\ee
as in \cite{cq} for example. Now in \cite{ne} one writes
\be
\oint_{B_j}d\Omega_n=\Omega_{nj}=-2\pi iRes\,z^{-n}d\omega_j=2\pi i\sigma_{jn}
\label{EWW}
\ee
and this is correct.  To see this go to the derivation in \cite{sc} where
one shows by the Riemann bilinear relations that
\be
\oint_{B_j}d\Omega_n=\Omega_{nj}=\sum\left(\oint_{A_i}d\omega_j
\oint_{B_i}d\Omega_n-\oint_{B_i}d\omega_j\oint_{A_i}d\Omega_n\right)=
\label{EXX}
\ee
$$=2\pi iRes(\omega_jd\Omega_n=2\pi iRes\left(-\sum\sigma_{jm}\frac
{z^m}{m}\right)\cdot\left(-nz^{-n-1}-\sum q_{mn}z^{m-1}\right)dz
=2\pi i\sigma_{jn}$$
Note also at this point that since the sum of residues of a meromorphic
differential is zero, one can write
$$
Res\,\Omega_md\Omega_n= q_{nm}-q_{mn}=0=$$
\be
=Res\left[z^{-m}-\sum_1^{\infty}q_{pm}\frac{z^p}{p}\right]
\cdot\left[-nz^{-n-1}-\sum_1^{\infty}q_{rn}z^{rn}z^{r-1}\right]
\label{ES}
\ee
showing that $q_{mn}=q_{nm}$.
Further we remark that $Res\,d(\omega_jd\Omega_n)=0=Res\,d\omega_j\Omega_n+
Res\,\omega_jd\Omega_n$.
\\[3mm]\indent
Now to deal with (\ref{EU}) one can write
\be
\Theta\left({\cal A}(P_{\infty})-\sum\frac{z^j}{j}
\frac{(\Omega_{jk})}{2\pi i}+\tilde{z}
\right)=\Theta\left({\cal A}(P)-\int_{P_{\infty}}^P(d\omega_k)-
\sum\frac{z^j}{j}\frac{(\Omega_{jk})}{2\pi i}+\tilde{z}\right)
\label{EV}
\ee
Since $\Omega_{jk}=2\pi i\sigma_{kj}$ and $-\int_{P_{\infty}}^Pd\omega_k
=-\omega_k=\sum\sigma_{km}(z^m/m)$ we have
a cancellation $\sum\sigma_{km}(z^m/m)-
\sum\sigma_{kj}(z^j/j)=0$ for $1\leq k\leq g$.  It follows that
\be
\Theta\left({\cal A}(P_{\infty})-\sum\frac{z^j}{j}(\Omega_{jk})+
\tilde{z}\right)=\Theta({\cal A}(P)+\tilde{z})
\label{EW}
\ee
and
\be
\frac{A}{B}=\frac{\Theta({\cal A}(P)+z_0)}{\Theta({\cal A}(P_{\infty})+z_0)}
\label{EX}
\ee
This will then lead to $d_k=d_k(\alpha)$.
Indeed (\ref{EB}) becomes now (after cancellation of the $t_j$ terms)
\be
0=\frac{1}{2}\sum_{k,l=1}^{\infty}q_{kl}\frac{z^{k+l}}{kl}-i\sum_1^g
\sum_1^{\infty}\alpha_j\sigma_{jk}\frac{z^k}{k}-\sum_1^{\infty}d_k\frac{z^k}
{k}+log\left(\frac{A}{B}\right)
\label{QW}
\ee
Writing ${\cal A}(P)={\cal A}(P_{\infty})+(\int_{P_{\infty}}^Pd\omega_j)$
and referring to \cite{dc} one has for $\check{z}={\cal A}(P_{\infty})
+z_0$
\be
log\Theta\left((\int_{P_{\infty}}^Pd\omega_j)+\check{z}\right)-log
\Theta(\check{z})=-\sum_{j=1}^g\sum_1^{\infty}(log\Theta)_j(\check{z})
\sigma_{jm}\frac{z^m}{m}+\cdots
\label{QV}
\ee
Putting this in (\ref{QW}) yields $d_k=d_k(\alpha)$ (cf. \cite{al,cq} for
another version of this).
\\[3mm]\indent
For perspective however let us make now a few background observations.  First
we refer first to \cite{ch} where it
is proved that $F_{mn}=F_{nm}$ in ${\cal B}_n=\lambda^n-\sum_1^{\infty}
(F_{nm}/m)\lambda^{-m}$ (the $F_{mn}$ being treated as algebraic symbols
with two indices).  Since near the point at infinity we have $\Omega_n\sim
\lambda^n-\sum_1^{\infty}(q_{mn}/m)\lambda^{-m}$ the same sort of proof
by residues is suggested ($F_{mn}=-Res_{\lambda}[{\cal B}_nd\lambda^m]$)
but we recall that ${\cal B}_n=\lambda^n_{+}$ so there is an underlying
$\lambda$ for all ${\cal B}_n$ which makes the proof possible.  Here this
is not a priori evident.  For example $(\bullet\spadesuit\bullet)\,\,
p=\lambda-\sum_1^{\infty}(H_j/j)\lambda^{-j}$ corresponds to $P=\lambda
+\sum_1^{\infty}P_{j+1}\lambda^{-j}$ in \cite{ch} with $P_{j+1}=F_{1j}/j$
(i.e. $H_j\sim -F_{1j}$) and the ``inverse" is $\lambda=P+\sum_1^{\infty}
U_{n+1}P^{-n}$ (arising from a Lax operator $L$ via dKP)  The corresponding
inverse for $(\bullet\spadesuit\bullet)$ then characterizes $\lambda$ in terms
of $p$ but one does not 
expect $\Omega_n\sim\lambda^n_{+}$.  The matter is somewhat
subtle.  Indeed the BA function is defined from the Riemann surface via
$d\Omega_n,\,\,d\omega_j$, and normalizations.  It then produces a unique
asymptotic expansion at $\infty$ which characterizes $\psi$ near $\infty$ in
terms of $\lambda$ and hence must characterize the $d\Omega_n$ and $d
\omega_j$ asymptotically.  Moreover the normalizations must be built into
these expansions since they were used in determining $\psi$.  But the 
formal algebraic determination of ${\cal B}_n$ via $\lambda^n_{+}$
is a consequence of relating the $d\Omega_n$ to operators $L_n=
L^n_{+}$ as in \cite{kt}.  Applying this reasoning it is suggested that
$(\bullet\clubsuit\bullet)$ is valid as a consequence of the BA
function linking the differentials and the asymptotic expansions.
Note that it is not stated that $q_{ij}=F_{ji}$ when $F_{ij}\sim
\partial_i\partial_jF$, and this point will be clarified below.  When
slow variables $T_n$ are introduced the moduli $h_k$ will be functions of
the $T_n$ ($h_k\sim$ branch points for hyperelliptic curves).  This means
that a priori the normalized differentials $d\Omega_n$ and $d\omega_j$
can depend on the $T_n$, along with $B_{ij}$, etc.  The construction
of $d{\cal S}=\sum T_nd\Omega_n+\sum a_jd\omega_j$ such that $\partial_n
d{\cal S}=d\Omega_n$ and $\partial_na_j=0$ is somewhat subtle and we will
look at this below following \cite{co,ib}.
Let us also say a few words about the construction (\ref{delta}), where
we recall the formulas (\ref{GC}) for example for the times $T_j$.  In the
present situation we simply think of $d{\cal S}=QdE$ in which case 
(\ref{GC}) becomes $T_j=(1/j)Res_{\infty}\lambda^{-j}d{\cal S}$.  Note
here from $d{\cal S}=\sum T_nd\Omega_n+\sum a_jd\omega_j$ with
$d\Omega_n=[n\lambda^{n-1}+O(\lambda^{-2})]d\lambda$ one obtains immediately
$Res_{\infty}(1/m)\lambda^{-m}d{\cal S}=T_m$.  As for the formula $\partial_n
F=Res_{\infty}\lambda^nd{\cal S}=-Res_0z^{-n}d{\cal S}$ we recall first the
form of $S$ in dKP, namely from Section 4.1, $S=\sum T_n\lambda^n+
\sum_1^{\infty}S_{j+1}\lambda^{-j}$ where in fact $S_{j+1}=
-(1/j)\partial_jF$ or simply $S_{j+1}\sim-(1/j)F_j$ for $F_j$ an algebraic
symbol
(cf. (\ref{YEEE}) and \cite{ch}).  Then in the dKP
case $dS=(\sum nT_n\lambda^{n-1}+\sum_1^{\infty}\partial_jF\lambda^{-j-1})
d\lambda$ so $Res_{\infty}\lambda^ndS=\partial_nF\sim F_n$ is immediate.  The
present situation 
with $d{\cal S}=\sum T_nd\Omega_n+\sum a_jd\omega_j,\,\,d\Omega_n=[n\lambda^
{n-1}+\sum q_{mn}\lambda^{-m-1}]d\lambda$, and $d\omega_j=\sum
\sigma_{jm}\lambda^{-m-1}d\lambda$ gives $Res\,\lambda^kd{\cal S}=
\sum T_nq_{kn}+\sum a_j\sigma_{jk}$ and this
is better treated indirectly as indicated below; note this suggests
$F_{km}=q_{km}$ (cf. \cite{cq}).
\\[3mm]\indent
Let us make some comments around the development in \cite{ib} (cf. also
\cite{co}).  For situations arising in Seiberg-Witten (SW) theory it is 
desired to find $d{\cal S}=\sum T_nd\Omega_n+\sum a_jd\omega_j$ with
$\partial_nd{\cal S}=d\Omega_n$ and $\partial_na_j=0$ so that $(T_n,a_j)$ can
be taken as independent variables with moduli $h_k=h_k(T,a)\,\,
(1\leq k\leq K$ - note generally $K>g$).  Here the moduli automatically 
depend on the slow variables $T_n$ and hence a priori so do the differentials.
First one notes that differentials $d\hat{\Omega}_n$, having the same
asymptotic properties as the $d\Omega_n$ (but unnormalized), can be selected
via $d\hat{\Omega}_n
=d[\lambda^n-
\sum_1^{\infty}\alpha_{in}(\lambda^{-i}/i)]$ with $\alpha_{in}$ ``arbitrary"
and $T$ independent.  Then we write $d\omega_j=-\sum_1^{\infty}
\sigma_{jm}z^{m-1}dz=\sum_{jm}\lambda^{-m-1}d\lambda$ where the 
$\sigma_{jm}$ depend on $h=(h_k)$ (and thence on $T$).  Now a general
desideratum in SW theory is that $\partial d{\cal S}/\partial h_k=
\sum\beta_{jk}d\omega_j$ and to achieve this one picks first the 
$d\hat{\Omega}_n$ to have this property.  Thus the $\alpha_{in}$ become 
functions of $h_k$ (hence of $T$) and we can specify $\partial d\hat
{\Omega}_n/\partial h_k=\sum_1^g\sigma^n_{ki}d\omega_i$ (for essentially
arbitrary $\sigma^n_{ki}$ - modulo the production of analytic functions,
which seems to allow enough flexibility here).  Note $\partial d\hat
{\Omega}_n/\partial h_k=[\sum_1^g(\partial\alpha_{in}\partial h_k)
\lambda^{-i-1}]d\lambda$ which can be written as a linear combination of the
$d\omega_j$ for each $n$ via 
\be
\frac{\partial\alpha_{in}}{\partial h_k}=\sum_j\sigma^n_{kj}\sigma_{ji}
\label{EH}
\ee
This defines $\partial\alpha_{in}/\partial h_k$ and hence the 
$\alpha_{in}$ via the $\sigma^n_{kj}$; a priori any (reasonable)
collection of $\sigma^n_{kj}$ should do.  To relate the $d\hat{\Omega}_n$
to $d\Omega_n$ we write then
\be
d\hat{\Omega}_n=d\Omega_n+\sum_1^gc^n_id\omega_i;\,\,c^n_i=\oint_{A_i}d\hat
{\Omega}_n
\label{EI}
\ee
Now impose the requirement $\partial_nd{\cal S}=d\Omega_n$ while
assuming $d{\cal S}=\sum u_m(T)d\hat{\Omega}_m$ (which guarantees that
$\partial d{\cal S}/\partial h_k=\sum\beta_{jk}d\omega_j$).  This leads to
\be
\partial_nd{\cal S}=\sum\left[\frac{\partial u_m}{\partial T_n}d\hat{\Omega}_m
+u_m\sum\frac{\partial d\hat{\Omega}_m}{\partial h_k}\partial_nh_k\right]=
\label{EJ}
\ee
$$=\sum\left[\partial_nu_md\hat{\Omega}_m+u_m\sum\partial_nh_k\sum\sigma^m_
{ki}d\omega_i\right]=d\Omega_n=d\hat{\Omega}_n-\sum c^nid\omega_i$$
This implies
\be
\partial_nu_m=\delta_{mn}\Rightarrow u_m=T_m;
\label{EK}
\ee
$$\sum\partial_nh_k\left(\sum T_m\sigma^m_{ki}\right)=\sum\partial_n
h_k\hat{\sigma}_{ki}=-c^n_i$$
where $\hat{\sigma}_{ki}=\sum_mT_m\sigma^m_{ki}$.  The equation
$(\spadesuit\spadesuit\spadesuit)\,\,\sum\partial_nh_k\hat{\sigma}_{ki}=
-c^n_i$ here represents the Whitham dynamics for the moduli $h_k$.  We note
that the $\sigma_{jm}(h)$ are inflexible but the terms $\sigma^n_{kj}$ are 
flexible; they determine the $c^n_i$ from (\ref{EI}) and the $\hat{\sigma}_
{ki}$ as in (\ref{EK}).  The Whitham dynamics are then ``compatible" equations
in some sense - given a choice of $\sigma^n_{ki}$ they describe a compatible
variation of moduli; however
they may not be determining for $h_k$ unless $K=g$
with $\hat{\sigma}_{ki}$ an invertible square matrix.  Now write
\be
d{\cal S}=\sum T_md\Omega_m+\sum T_m\sum c^m_id\omega_i=
\label{EL}
\ee
$$=\sum T_md\Omega_m+\sum a_id\omega_i;\,\,a_i=\sum T_mc^m_i=
\oint_{A_i}d{\cal S}$$
and look at the requirement $\partial_na_i=0$ which would yield
$a_i$ as an independent variable with $h_k=h_k(T_n,z_i)$ along with
$\partial d{\cal S}/\partial a_i=d\omega_i$.  This situation is handled
in \cite{ib} via an assumption $K=g$ leading to determination of the
$\partial_nh_k$ via $(\clubsuit\clubsuit\clubsuit)\,\,\partial_nh_k
=-\sum c^n_i\hat{\sigma}_{ik}^{-1}$.  Additional calculation yields then
$(\partial h_k/\partial a_j)|_{T=c}=\hat{\sigma}_{jk}^{-1}(h)$.  One can
also examine matters however without making any assumption on $K$.  Thus
e.g. given (\ref{EL}), $\partial_na_j=0$ requires
\be
\partial_na_j=\partial_n\oint_{A_j}d{\cal S}=\oint_{A_j}d\Omega_n=0
\label{EM}
\ee
which is automatically true for this standard normalization of $d\Omega_n$.
Therefore without explicitly solving for $h_k(T_n,a_j)$ we can say
that this form prevails when the ``Whitham equations" $(\spadesuit
\spadesuit\spadesuit)$ for $h_k$ have a solution (perhaps not unique).
In these circumstances we will have also $\partial 
d{\cal S}/\partial a_i=d\omega_i$ and
\be
a_j^D=\oint_{B_j}d{\cal S}=\oint_{B_j}\left(\sum a_id\omega_i+\sum
T_nd\Omega_n\right)=\sum a_iB_{ij}+\sum T_n\Omega_{nj}
\label{EN}
\ee
\indent
Let us develop this a little further following \cite{ne}.  We will simply 
repeat some of the argument there since it is instructive and revealing.
We assume the $d\Omega_n$ and $d\omega_j$ are specified as before with
$d\Omega_n=(-nz^{-n-1}-\sum_1^{\infty}q_{mn}z^{m-1})dz$ and $d\omega_j=
-\sum_1^{\infty}\sigma_{jm}z^{m-1}dz$.  The standard Whitham equations
based on times $T_n$ and $a_j$ are (see Section 5)
\be
\partial_n d\Omega_m=\partial_md\Omega_n;\,\,\partial_nd\omega_j=\frac
{\partial d\Omega_n}{\partial a_j};\,\,\frac{\partial d\omega_j}{\partial a_i}=
\frac{\partial d\omega_i}{\partial a_j}
\label{EO}
\ee
This form automatically leads to a differential $d{\cal S}$ such that
\be
\partial_nd{\cal S}=d\Omega_n;\,\,\frac{\partial d{\cal S}}{\partial a_j}
=d\omega_j
\label{EP}
\ee
for which (\ref{EO}) describes compatibility conditions.  Then one is led
(automatically) to construct a function $F$ satisfying (\ref{delta}).
The consistency of this stipulation (\ref{delta}) is ensured by (\ref{EO})
and the Riemann bilinear relations.  For example to prove $\partial_n(\partial
F/\partial a_i)=\partial(\partial_nF)/\partial a_i$ using the formulas of (\ref
{delta}) one can write ($\Omega_n=\int^Pd\Omega_n$)
\be
\frac{\partial}{\partial T_n}\left(\frac{1}{2\pi i}\oint_{B_i}d{\cal S}\right)
=\frac{1}{2\pi i}\oint_{B_i}d\Omega_n=
\label{EQ}
\ee
$$=-\frac{1}{2\pi i}\sum_1^g\left\{\oint_{A_j}d\Omega_n\oint_{B_j}d\omega_i
-\oint_{A_j}d\omega_i\oint_{B_j}d\Omega_n\right\}=$$
$$=-Res_{\infty}\omega_nd\omega_i=-Res\,z^{-n}d\omega_i$$
(cf. (\ref{CC}) and (\ref{EXX})).  
Finally one shows that $\partial_n\partial_mF=
\partial_m\partial_nF$ via
$$
-\partial_nRes\,z^{-m}d{\cal S}=-Res\,z^{-m}d\Omega_n=q_{mn}=$$
\be
=-Res\,z^{-m}\left[-nz^{-n-1}-\sum_1^{\infty}q_{pn}z^{p-1}\right]dz
\label{ER}
\ee
whereas $\partial_mRes\,z^{-n}d{\cal S}=q_{nm}$.  
From the relation (\ref{delta}) we see also that the local behavior of 
$d{\cal S}$ can be described in terms of $F$ via
\be
d{\cal S}=-\sum nT_nz^{-n-1}-\sum_1^{\infty}\partial_nF\,z^{n-1}
\label{ET}
\ee

\end{document}